\documentclass{jpp}
\usepackage{graphicx}
\usepackage[utf8]{inputenc}
\usepackage[T1]{fontenc}
\usepackage{amsmath}

\usepackage{bm}
\usepackage{amsmath}
\usepackage{xcolor}
\newcommand{\partder}[2]{\frac{\partial #1}{\partial #2}}

\usepackage{cleveref}
\crefname{equation}{}{}
\crefname{figure}{figure}{figures}
\usepackage{subcaption}
\usepackage[normalem]{ulem}

\shorttitle{}
\shortauthor{}

\title{Gradient-based optimization of 3D MHD equilibria}
\shorttitle{Gradient-based optimization of 3D MHD equilibria}
\author{Elizabeth J. Paul\aff{1}
  \corresp{\email{epaul@princeton.edu}}, Matt Landreman\aff{2}, Thomas Antonsen, Jr.\aff{2}}
 \shortauthor{Elizabeth J. Paul, Matt Landreman, Thomas Antonsen, Jr.}

\affiliation{\aff{1}Department of Astrophysical Sciences, Princeton University,
Princeton, NJ 08544, USA 
\aff{2} Institute for Research in Electronics and Applied Physics, University of Maryland,
College Park, MD 20742, USA}

\begin{document}

\maketitle

\begin{abstract}
Using recently developed adjoint methods for computing the shape derivatives of functions that depend on MHD equilibria \citep{Antonsen2019,Paul2020}, we present the first example of analytic gradient-based optimization of fixed-boundary stellarator equilibria. We take advantage of gradient information to optimize figures of merit of relevance for stellarator design, including the rotational transform, magnetic well, and quasisymmetry near the axis. With the application of the adjoint method, we reduce the number of equilibrium evaluations by the dimension of the optimization space ($\sim50-500$) in comparison with a finite-difference gradient-based method. We discuss regularization objectives of relevance for fixed-boundary optimization, including a novel method that prevents self-intersection of the plasma boundary. We present several optimized equilibria, including a vacuum field with very low magnetic shear throughout the volume. 
\end{abstract}

\section{Introduction}

The improved confinement of modern stellarators is largely attributed to the numerical optimization of the magnetic field. With the technique pioneered by N\"{u}hrenberg and Zille \citep{Nuhrenberg1988}, the shape of the plasma boundary, $S_P$, is optimized for certain properties of the magnetohydrodynamic (MHD) equilibrium,
\begin{subequations}
\begin{align}
    \left(\nabla \times \textbf{B}\right) \times \textbf{B} &= \mu_0\nabla p  &\text{in } V_P\\
    \textbf{B} \cdot \hat{\textbf{n}} &= 0 &\text{on } S_P, \label{eq:neumann}
\end{align}
\label{eq:MHD}
\end{subequations}
where $S_P = \partial V_P$, $p(\psi)$ is the prescribed pressure profile, and $\hat{\textbf{n}}$ is the unit normal. For the past three decades, stellarator optimization has largely proceeded with this fixed-boundary approach with codes such as STELLOPT \citep{Lazerson2020}
and ROSE \citep{Drevlak2018}, resulting in the W7-X \citep{Beidler1990}, HSX \citep{Anderson1995}, and NCSX \citep{Zarnstorff2001} configurations. We use the term ``fixed-boundary'' to describe this approach as the equilibrium is computed with a specified boundary, although the boundary is varied throughout the optimization. 

Although there has been significant success with this approach, there are several ways to improve the performance of fixed-boundary optimization. Specifically, the incorporation of derivative information could be transformative. Fixed-boundary optimization has previously relied on derivative-free methods -- such as genetic algorithms \citep{Miner2001}, differential evolution \citep{Mynick2002}, and Brent's algorithm \citep{Drevlak2018} -- or derivative-based methods with finite-difference gradients. While global derivative-free algorithms may prevent the optimization from terminating in local minima, they are only effective for smaller problems \citep{Nocedal2006}. Gradient-based optimization with finite-difference gradients requires excessive function evaluations in high-dimensional spaces and suffers from error that enters due to the choice of step size. If the step size is too small, the error is dominated by round-off error, and if it is too large it is dominated by nonlinearity \citep{Sauer2012}. Because of the requirement of excessive function evaluations and the unreliability of the gradient information, gradient-based optimization with finite-difference derivatives is not always effective. In this work, we present the first example of analytic gradient-based, fixed-boundary optimization of stellarator equilibria.

With a local gradient descent approach, each iteration reduces to a one-dimensional line search \citep{Nocedal2006}; thus, the further incorporation of derivative information eliminates restrictions on the size of the optimization space. There is some evidence from the machine learning community that overparameterization of the space can accelerate optimization \citep{Oymak2018}. Therefore, it is possible that increasing the Fourier resolution of the plasma boundary may similarly eliminate local minima.

There are several ways that this derivative information can be obtained. For sufficiently simple figures of merit, the objective can be analytically differentiated and implemented by hand. Alternatively, the derivatives can be obtained programmatically using automatic differentiation tools. When a given objective function depends on the solution of a set of equations, such as the MHD equilibrium equations \eqref{eq:MHD}, the derivatives can be obtained more efficiently using an adjoint method. With this technique, the solution of only one additional equation, known as the adjoint equation, is required. Once the adjoint solution is obtained, the derivative of a given objective can be obtained with respect to any optimization parameter, eliminating the need to solve a perturbed set of equations. In this way, the cost associated with obtaining a high-dimensional gradient is significantly reduced. In this work, we employ adjoint-based gradients of functions which depend on the MHD equilibrium equations. This adjoint method results from a generalized self-adjointness property of the MHD force operator \citep{Antonsen2019}. 
This technique has been demonstrated for computing the shape derivatives of several figures of merit relevant for stellarator design, including the magnetic well, rotational transform, and magnetic ripple \citep{Paul2020}. Each of these objective functions have been included in modern stellarator designs \citep{Beidler1990,Anderson1995,Zarnstorff2001,Henneberg2019}. We also employ analytic gradients for objectives that do not depend on the MHD equilibrium equations, such as the volume and properties of the surface curvatures.

There have been several recent applications of derivative information to other related problems in stellarator design. The FOCUS \citep{Zhu2018} and FOCUSADD \citep{Mcgreivy2021} codes optimize coil shape to be consistent with a given plasma boundary with gradients obtained from analytic and automatic differentiation methods, respectively. Our work is distinct from the FOCUS approach, as we use gradients to optimize properties of the equilibrium rather than using gradients to optimize coils in order to match the boundary of a given equilibrium. Adjoint methods have recently been applied to directly optimize coils for quasisymmetry near the magnetic axis in a vacuum field \citep{Giuliani2020}. In contrast, our approach can be applied to optimize equilibria with arbitrary pressure. Furthermore, we have developed adjoint methods for direct optimization of coil shapes for properties of an MHD equilibrium \citep{Antonsen2019,Paul2020}, although its application to optimization is not presented in this work. Adjoint methods have also been used to optimize the local magnetic field for neoclassical properties \citep{Paul2019} and the coil winding surface for properties of the current potential \citep{Paul2018}.

We discuss the new gradient-based fixed-boundary optimization tool in \S \ref{sec:overview}. In \S \ref{sec:regularization}, we present regularization terms for fixed-boundary optimization, including a constraint on the curvature of the boundary that prevents self-intersection. In \S \ref{sec:demonstration}, we present several optimization demonstrations, including obtaining a low magnetic shear configuration (\S \ref{sec:iota_target}), a configuration with a magnetic well (\S \ref{sec:well}), and a configuration with quasisymmetry near the magnetic axis (\S \ref{sec:ripple}). We conclude in \S \ref{sec:conclusions}.

\section{Overview of ALPOpt optimization tool}
\label{sec:overview}

As with the STELLOPT and ROSE codes, the ALPOpt\footnote{https://github.com/ejpaul/ALPOpt} tool optimizes the boundary of VMEC \citep{Hirshman1983} equilibria. The VMEC code obtains solutions of \eqref{eq:MHD} under the assumption of nested toroidal magnetic surfaces. The plasma boundary is described by a set of Fourier coefficients of the cylindrical coordinates, $\{R_{m,n}^c,Z_{m,n}^s\}$,
\begin{subequations}
\begin{align}
    R(\theta,\phi) &= \sum_{m,n} R_{m,n}^c \cos(m \theta - n N_P \phi) \\
    Z(\theta,\phi) &= \sum_{m,n} Z_{m,n}^s \sin(m \theta - nN_P \phi),
\end{align}
\label{eq:rmnc_zmns}
\end{subequations}
where $\theta$ is a poloidal angle, $\phi$ is the cylindrical toroidal angle, and $N_P$ is the number of field periods. Therefore, the optimization space is taken to be the set of coefficients $\{R_{m,n}^c, Z_{m,n}^s\}$. The pressure profile $p(\psi)$ and another function of flux, either the rotational transform $\iota(\psi)$ or enclosed toroidal current $I_T(\psi)$, are prescribed and fixed. (For all of the examples in this work, $I_T(\psi)$ is fixed.) The optimization code interfaces with VMEC through python, and optimization is performed with the scipy\footnote{https://docs.scipy.org/doc/scipy/reference/generated/scipy.optimize.minimize.html} and NLOPT \citep{NLOPT} packages. 

The optimization tool takes advantage of the adjoint method for obtaining the shape gradient of MHD equilibria \citep{Antonsen2019,Paul2020}. 
Here the perturbation to the magnetic field due to a perturbation of the boundary is expressed as,
\begin{align}
\delta \textbf{B}_1 = \nabla \times \left(\bm{\xi}_1 \times \textbf{B}\right) - \delta \iota_1(\psi)\nabla \psi \times \nabla \phi,
\end{align}
where the displacement vector satisfies $\bm{\xi}_1 \cdot \hat{\textbf{n}} = \delta \textbf{x} \cdot \hat{\textbf{n}}$ on the boundary for a given normal perturbation to the surface and $\delta \iota_1(\psi)$ is the perturbation to the rotational transform profile that may arise due to the constraint of fixed $I_T(\psi)$. This perturbed magnetic field can be related to an \textit{adjoint} perturbed magnetic field, $\delta \textbf{B}_2$, through a generalized self-adjointness relation,
\begin{multline}
    \int_{V_P} d^3 x \, \left(\bm{\xi}_1 \cdot  \textbf{F}_2   - \bm{\xi}_2 \cdot  \textbf{F}_1 \right)
    - 2\pi \int_{V_P} d \psi \, \left( \delta I_{T,2}(\psi) \delta \iota_{1}(\psi) - \delta I_{T,1}(\psi) \delta \iota_{2}(\psi) \right) \\
    - \frac{1}{\mu_0} \int_{S_P} d^2 x \, \hat{\textbf{n}} \cdot \left( \bm{\xi}_2 \delta \textbf{B}_1 - \bm{\xi}_1 \delta \textbf{B}_2  \right) \cdot \textbf{B} = 0,
    \label{eq:fixed_boundary}
\end{multline}
where the linearized force operator is,
\begin{align}
     \textbf{F}_{1,2} = \frac{\left(\nabla \times \delta \textbf{B}_{1,2} \right) \times \textbf{B}}{\mu_0} + \frac{\left(\nabla \times \textbf{B} \right) \times \delta \textbf{B}_{1,2}}{\mu_0} + \nabla \left(\bm{\xi}_{1,2} \cdot \nabla p \right).
     \label{eq:force_operator}
\end{align}
The adjoint approach is as follows: rather than directly computing the perturbed magnetic field $\delta \textbf{B}_1$ by perturbing the boundary of a fixed-boundary equilibrium or solving a set of linearized equilibrium equations, the adjoint magnetic field is computed. The adjoint magnetic field has no perturbation to the boundary ($\bm{\xi}_2 \cdot \hat{\textbf{n}} = 0$) but may have a perturbation to the toroidal current profile ($\delta I_{T,2}(\psi)$) or a bulk force perturbation ($\textbf{F}_2$). In this work, we will consider figures of merit whose derivatives can be computed with a perturbation to the toroidal current profile or a bulk force perturbation which takes the form of the gradient of a scalar function of flux or the divergence of an anisotropic pressure tensor.

Rather than consider a set of linearized equations, we add a small perturbation to a nonlinear VMEC equilibrium in the form of a perturbation to the pressure profile, toroidal current profile, or an anisotropic pressure tensor. In the case of the addition of a pressure tensor, the ANIMEC \citep{Cooper19923d} code is used to evaluate the adjoint equilibrium. The resulting shape gradient, $\mathcal{G}$, of a given objective function $f$, is defined through,
\begin{align}
    \delta f(S_P;\delta \textbf{x}) = \int_{S_P} d^2 x \, \delta \textbf{x} \cdot \hat{\textbf{n}} \mathcal{G}
    \label{eq:shape gradient}.
\end{align}
Here $\delta f(S_P;\delta \textbf{x})$ is the shape derivative of $f$ with respect to a normal perturbation of the surface, $\delta \textbf{x} \cdot \hat{\textbf{n}}$. Given the shape gradient, which quantifies the local sensitivity to normal perturbations of the surface, the derivatives with respect to the parameters $\Omega = \{R_{m,n}^c,Z_{m,n}^s\}$ are computed,
\begin{align}
    \partder{f}{\Omega} &= \int_{S_P} d^2 x \, \partder{\textbf{x}}{\Omega} \cdot \hat{\textbf{n}} \mathcal{G}.
\end{align}

\subsection{Managing code failures}

The equilibrium code may return with an error for a given plasma boundary. There are many reasons for these failures, such as the flux coordinate Jacobian becoming ill-conditioned or the number of iterations exceeding the maximum prescribed value. When such a failure is experienced during the optimization, the objective function is set to an arbitrarily large value (e.g., $10^{12}$) to enforce an effective constraint. The approach of assigning a very large fictitious objective value at unevaluable points is common in the optimization literature \citep{Rasheed1997,Emmerich2002} and is employed in the STELLOPT code \citep{Lazerson2020}.

As encountering such unevaluable points makes the parameter space non-smooth, it is prudent to try to avoid code failures by placing additional constraints on the parameter space. Such a technique has been employed in the optimization of aircraft by checking that simulation outputs match physical model assumptions, such as the drag coefficient being non-negative \citep{Gelsey1995,Gelsey1998}. In \S \ref{sec:regularization}, we present constraints which prevent the surface from self-intersecting or from obtaining large curvature, leading to an ill-conditioned Jacobian and code failure. 
Even in the presence of such constraints, unevaluable points may still be present. Here the objective function becomes discontinuous, which is problematic for optimizers which assume function continuity. In this work we employ the BFGS quasi-Newton method with an Armijo-Wolfe line search, which assumes $C^2$ continuity of the objective function \citep{Nocedal2006}. Although convergence of the BFGS algorith is not guaranteed for non-smooth problems, it has been observed that a reasonable approximation to the optimum is most often achieved \citep{Lemarechal1982}. With the standard Armijo-Wolfe line search method, the gradient need not be evaluated unless the objective function satisfies the sufficient decrease criterion. Thus the gradient does not need to be evaluated at unevaluable points. In this way, as long as the line search never returns a point where the objective is not differentiable, the BFGS method is well defined. While there are specialized quasi-Newton methods for non-smooth objective functions \citep{Lewis2009}, we have obtained acceptable results with a standard BFGS method.




\section{Regularization terms}
\label{sec:regularization}

\subsection{Preventing surface self-intersection}
\label{sec:global_curvature}

We now describe a constraint which prevents self-intersection of the plasma boundary. Given a surface described by the cylindrical coordinates $R(\theta,\phi)$ and $Z(\theta,\phi)$, self-intersection of the boundary may occur if either:
\begin{align*}
&\text{(a) $R(\theta,\phi)< 0$ at any point, or} \\
&\text{(b) the planar curve $\textbf{x}_{\phi_0}(\theta) = \{ R(\theta,\phi_0), Z(\theta,\phi_0)\}$ is self-intersecting for any $\phi_0$.}
\end{align*}
Condition (a) can be avoided using a penalty objective of the form,
\begin{align}
    f_R(S_P) = \frac{1}{A_P} \int_{S_P} d^2 x \, \exp \left(- \frac{\left(R - R_{\text{min}}\right)}{w_R}  \right),
\end{align}
where $w_R$ is a weight which sets the gradient length scale of the objective, $R_{\text{min}}$ is the minimum allowable major radius, and $A_P$ is the area of $S_P$.

\begin{figure}
    \centering
    \begin{subfigure}{0.33\textwidth}
    \includegraphics[trim=9cm 5cm 
    15cm 2cm,clip,width=1.0
    \textwidth]{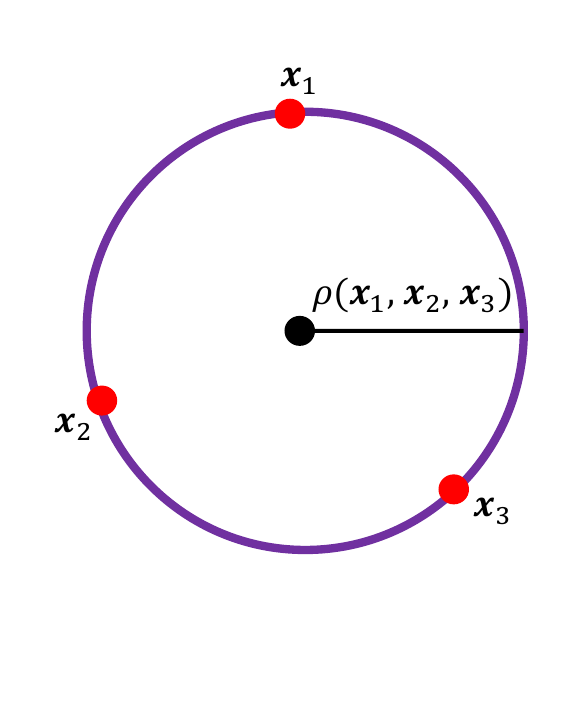}    
    \caption{}
    \end{subfigure}
    \begin{subfigure}{0.66\textwidth}
    \includegraphics[trim=4cm 3cm 
    4cm 0cm,clip,width=1.0
    \textwidth]{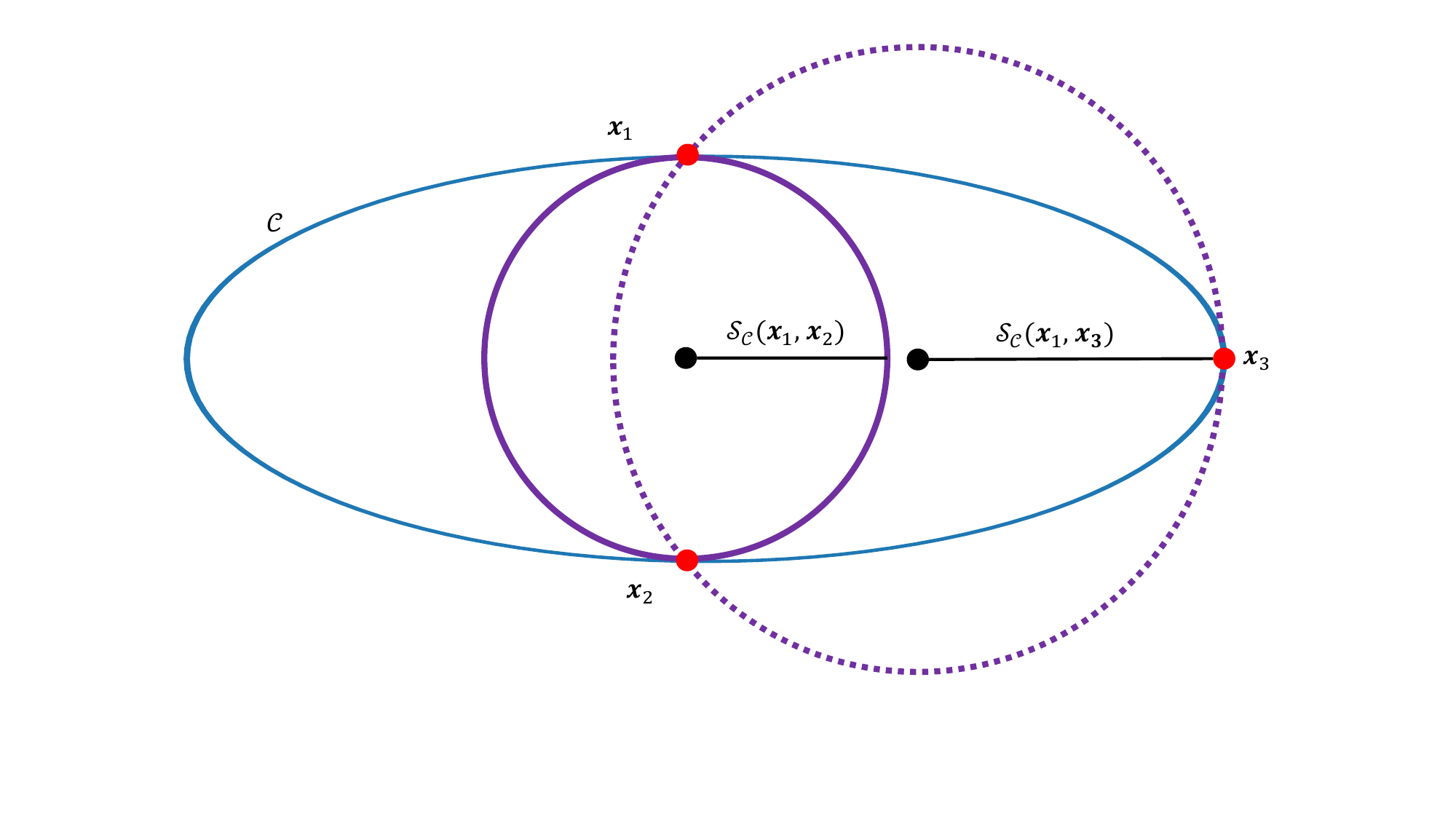}   
    \caption{}
    \end{subfigure}
    \caption{(a) The circumcircle containing $\textbf{x}_1$, $\textbf{x}_2$, and $\textbf{x}_3$ with radius $\rho(\textbf{x}_1,\textbf{x}_2,\textbf{x}_3)$. (b) The points $\textbf{x}_1$, $\textbf{x}_2$, and $\textbf{x}_3$ lie on a closed elliptical curve $\mathcal{C}$ (blue). The point tangent radius for points $\textbf{x}_1$ and $\textbf{x}_2$ and points $\textbf{x}_1$ and $\textbf{x}_3$ are displayed. The global radius of curvature at $\textbf{x}_1$ is $\mathcal{S}_{\mathcal{C}}(\textbf{x}_1,\textbf{x}_2)$.}
    \label{fig:pointtangent}
\end{figure}

Condition (b) can be avoided by introducing a constraint on the \textit{global radius of curvature} \citep{Gonzalez1999,Walker2016} of each of the curves, $\textbf{x}_{\phi_0}(\theta)$, which will now be defined. The radius of the unique circumcircle containing any three points $\textbf{x}_1$, $\textbf{x}_2$, and $\textbf{x}_3$ (Figure \ref{fig:pointtangent}), can be computed from,
\begin{align}
    \rho(\textbf{x}_1,\textbf{x}_2,\textbf{x}_3) = \frac{|\textbf{x}_1 - \textbf{x}_2||\textbf{x}_1 - \textbf{x}_3||\textbf{x}_2-\textbf{x}_3|}{4 \mathcal{A}(\textbf{x}_1,\textbf{x}_2,\textbf{x}_3)},
\end{align}
where $\mathcal{A}(\textbf{x}_1,\textbf{x}_2,\textbf{x}_3)$ is the area of a triangle with vertices $\textbf{x}_1$, $\textbf{x}_2$, and $\textbf{x}_3$. Under the assumption that these points lie on a non-self-intersecting smooth curve, $\mathcal{C}$ such that $\textbf{x}_1 = \textbf{x}(l_1)$, $\textbf{x}_2 = \textbf{x}(l_2)$, and $\textbf{x}_3 = \textbf{x}(l_3)$, then the radius of the circumcircle satisfies  $\rho(\textbf{x}(l_1),\textbf{x}(l_2),\textbf{x}(l_2)) \le \rho(\textbf{x}(l_1),\textbf{x}(l_2),\textbf{x}(l_3))$. This limiting case can be computed from,
\begin{align}
    \rho(\textbf{x}(l_1),\textbf{x}(l_2),\textbf{x}(l_2)) = \lim_{l_2 \rightarrow l_3} \rho(\textbf{x}(l_1),\textbf{x}(l_2),\textbf{x}(l_3)) = \frac{|\textbf{x}(l_1)-\textbf{x}(l_2)|}{2\sqrt{1 - \left(\hat{\textbf{t}}(\textbf{x}(l_2)) \cdot \frac{\textbf{x}(l_1)-\textbf{x}(l_2)}{|\textbf{x}(l_1)-\textbf{x}(l_2)|} \right)^2}},
\end{align}
where $\hat{\textbf{t}}$ is the unit tangent. This is the radius of the circle that passes through $\textbf{x}(l_1)$ and is tangent to $\mathcal{C}$ at $\textbf{x}(l_2)$, which we define as the self-contact function between these points,
\begin{align}
    \mathcal{S}_{\mathcal{C}}(\textbf{x}_1,\textbf{x}_2) = \frac{|\textbf{x}_1 - \textbf{x}_2|}{2 \sqrt{1-\left(\hat{\textbf{t}}(\textbf{x}_2) \cdot \frac{\left(\textbf{x}_1-\textbf{x}_2 \right)}{|\textbf{x}_1 - \textbf{x}_2|}\right)^2}},
\end{align}
the radius of the so-called point-tangent circle (Figure \ref{fig:pointtangent}). 

In the limit that $l_2, l_3 \rightarrow l_1$, we obtain
\begin{align}
    \lim_{l_2,l_3 \rightarrow l_1} \rho(\textbf{x}(l_1),\textbf{x}(l_2),\textbf{x}(l_3)) = \rho_{\mathcal{C}}(\textbf{x}(l_1)),
\end{align}
where $\rho_{\mathcal{C}}(\textbf{x}(l_1))$ is the local radius of curvature of $\mathcal{C}$ at $\textbf{x}(l_1)$. We define the global radius of curvature as,
\begin{align}
    \rho_{G}(\textbf{x}_1) = \min_{\textbf{x}_2,\textbf{x}_3 \in \mathcal{C}} \rho(\textbf{x}_1,\textbf{x}_2,\textbf{x}_3) = \min_{\textbf{x}_2 \in \mathcal{C}} \mathcal{S}_{\mathcal{C}}.
    \label{eq:rho_G}
\end{align}
This can be thought of as a generalization of the local radius of curvature, as we have the inequality $0 \le \rho_G(\textbf{x}) \le \rho_{\mathcal{C}}(\textbf{x})$. At a given point, the minimizer in \eqref{eq:rho_G} will either be a point where $(\textbf{x}_1 - \textbf{x}_2) \cdot \hat{\textbf{t}}(\textbf{x}_2) = 0$ or $\rho_{\mathcal{C}}(\textbf{x}_2) = \mathcal{S}_{\mathcal{C}}(\textbf{x}_1,\textbf{x}_2)$ \citep{Smutny2004}. In other words, at a given point $\textbf{x}_1$, the point on the curve that has the smallest self-contact function containing $\textbf{x}_1$ will be where the local radius of curvature is equal to the point-tangent radius or the displacement of the points $(\textbf{x}_1 - \textbf{x}_2)$ is orthogonal to the curve at $\textbf{x}_2$.

If the inequality
\begin{align}
   \min_{\textbf{x}_1 \in \mathcal{C}} \rho_G(\textbf{x}_1) = \min_{\textbf{x}_1,\textbf{x}_2 \in \mathcal{C}} \mathcal{S}_{\mathcal{C}}(\textbf{x}_1,\textbf{x}_2) \ge \mathcal{R}
   \label{eq:global_curvature_constraint}
\end{align}
is satisfied, this implies that the curve can be ``thickened'' to a tube without self-contact surrounding $\mathcal{C}$ such that at any given point, the cross-section of the tube is a circle in the plane perpendicular to the local tangent vector of radius $\mathcal{R}$ \citep{Gonzalez1999}.  For this reason, the concept of the global radius of curvature has been employed for the shape optimization of finite-thickness knots \citep{Gonzalez1999,Carlen2005,Walker2016}. 

We enforce the inequality constraint \eqref{eq:global_curvature_constraint} to prevent the self-intersection of every planar curve $\textbf{x}_{\phi_0}(\theta)$ with a penalty function of the form,
\begin{align}
    f_{\mathcal{S}} = \frac{1}{A_P}\int_0^{2\pi} d \phi \int_0^{2\pi} d\theta \, \left|\partder{\textbf{r}}{\theta} \times \partder{\textbf{r}}{\phi}\right|\int_0^{2\pi} d \theta' \,   \exp\left(-\frac{\left(\mathcal{S}_{\mathcal{C}}(\textbf{x}_{\phi}(\theta),\textbf{x}_{\phi}(\theta')) - \mathcal{R}\right)}{w_{\mathcal{S}}}\right),
\end{align}
where $\mathcal{R}$ is the minimum allowable global radius of curvature and $w_{\mathcal{S}}$ is a weight function. In addition to preventing self-intersection of the boundary, this objective improves the regularity of the surface by reducing the curvature of the poloidal cross-sections. A demonstration of this penalty function is presented in \S \ref{sec:constrained_volume}.

\subsection{Surface curvature constraint}

While constraining the self-contact function reduces the curvature of the cross-sections of the plasma boundary, we include additional terms which improve the regularity of the plasma boundary. First, we place an effective constraint on the magnitude of the principal curvatures,
\begin{align}
    |\kappa_1| &\le \kappa_{\max,1} &   |\kappa_2| \le \kappa_{\max,2},
\end{align}
by including a penalty function of the form, 
\begin{align}
    f_{\kappa} = \frac{\int_{S_P} d^2 x \, \left(e^{(|\kappa_1|^2-\kappa_{\max,1}^2)/w_k^2} 
    + e^{(|\kappa_2|^2-\kappa_{\max,2}^2)/w_k^2}\right)}{A_P}.
    \label{eq:fkappa}
\end{align}
A similar constraint on the principal curvatures is employed in the ROSE code \citep{Drevlak2018}. We assume the convention that $\kappa_1 \le \kappa_2$ and $\kappa_{1,2}<0$ indicates concavity.

We additionally add a penalty on the smallest principal curvature,
\begin{align}
    f_{\overline{\kappa}} = \frac{\int_{S_P} d^2 x \, \kappa_1^2}{A_P}.
\end{align}
As large concavity is associated with coil complexity \citep{Paul2018}, this regularization becomes critical in the two-staged optimization approach. Also, regularity of the boundary improves the convergence of the VMEC code. If the boundary is not close to being a star domain, indicating that there exists a coordinate axis such that the line segment connecting the axis and any point on the boundary is contained within the boundary, then the solver may fail to initialize a guess for the magnetic axis. For highly-shaped boundaries, the coordinate surfaces may begin to overlap. Thus, the inclusion of these regularization terms prevents code failures during optimization and improves the convergence toward the optimum. This advantage of curvature penalization is highlighted in the magnetic well optimization in \S \ref{sec:well}.

\subsection{Constrained volume optimization}
\label{sec:constrained_volume}

To demonstrate the procedure described in \S \ref{sec:global_curvature} for avoiding self-intersection, we define an objective function,
\begin{align}
    f(S_P) = V_P + \lambda_{\mathcal{S}} f_{\mathcal{S}} + \lambda_{R} f_R + \lambda_{\kappa} f_{\kappa},
    \label{eq:volume_objective}
\end{align}
where $V_P$ is the volume enclosed by the plasma boundary. For this example, we take $\lambda_{\mathcal{S}} = \lambda_{R} = 10^3$, $R_{\text{min}} = \mathcal{R} = 0.2$ m, $w_R = w_{\mathcal{S}} = 10^{-2}$, $\lambda_{\kappa} = 1$, $\kappa_{\text{max},1} = \kappa_{\text{max},2} = 10$ m$^{-1}$, and $w_k = 1$. Here the volume is computed using a surface integral upon application of the divergence theorem,
\begin{align}
    V_P &= \int d^3 x = \int_{S_P} d^2 x \, Z \hat{\textbf{n}} \cdot \hat{\textbf{Z}},
    \label{eq:volume_surface_integral}
\end{align}
to avoid discretizing the volume.


We begin with a stellarator whose boundary is defined by a rotating ellipse,
\begin{subequations}
\begin{align}
    R(\theta,\phi) = R_0 &+ 0.5a\left(\cos \left(\theta - N_{P} \phi\right) + \cos(\theta) \right) \nonumber \\
    &- 0.5 b \left(\cos\left(\theta - N_{P}\phi\right) -\cos(\theta)\right)
\end{align}
\begin{align}
    Z(\theta,\phi) = &0.5b\left(\sin(\theta) + \sin\left(\theta - N_{P}\phi\right) \right) \nonumber \\
    &+ 0.5 a \left(\sin(\theta) -\sin\left(\theta - N_{P} \phi\right)\right),
\end{align}
\label{eq:initial_volume}
\end{subequations}
where $a$ is the semi-major axis and $b$ is the semi-minor axis. For this example, we take $R_0 = 5$, $a =2$, $b= 1$, and $N_P=3$. We optimize $f(S_P)$ with respect to modes $m \le 3$ and $|n| \le 3$ using the BFGS algorithm provided by scipy. (This optimization algorithm is used for all demonstrations in this work.) 

\begin{figure}
\centering
    \begin{subfigure}{0.37\textwidth}
    \includegraphics[trim=0cm 3cm 
    7.5cm 3cm,clip,width=1.0
    \textwidth]{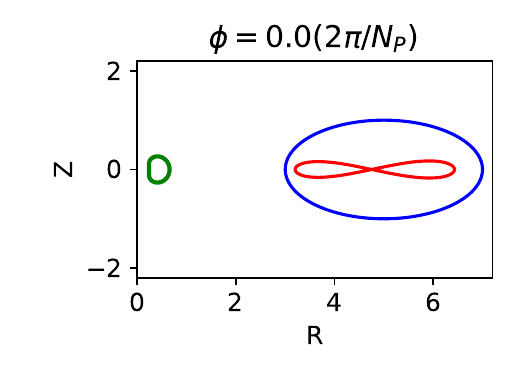}    
    \end{subfigure}
    \begin{subfigure}{0.62\textwidth}
    \includegraphics[trim=1.5cm 3cm 
    0cm 3cm,clip,width=1.0
    \textwidth]{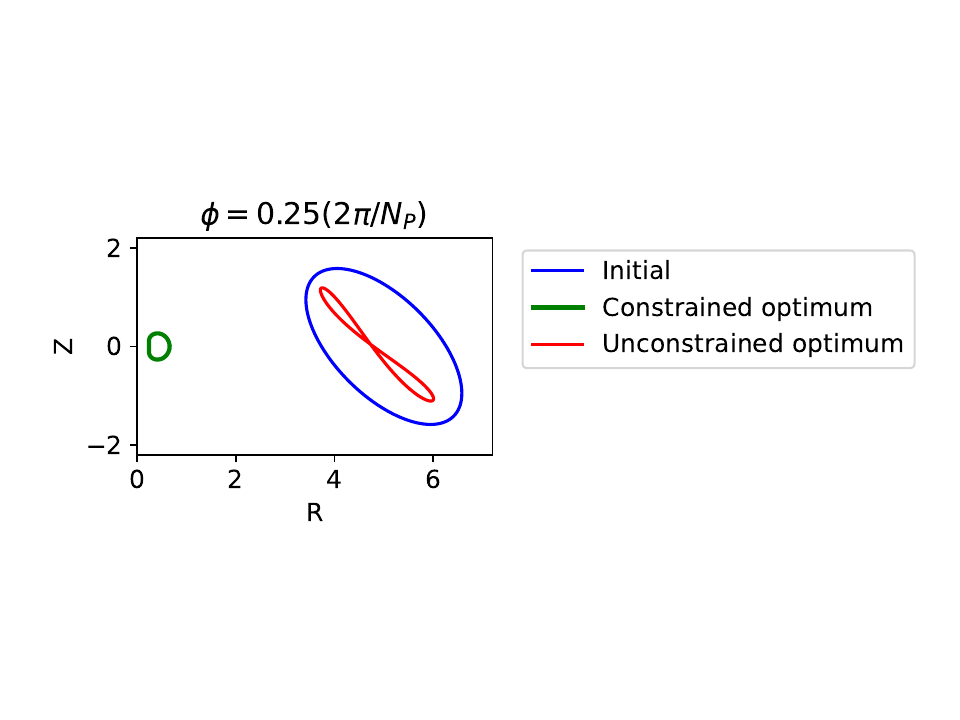}    
    \end{subfigure}
    \begin{subfigure}{0.37\textwidth}
    \includegraphics[trim=0cm 2cm 
    7.5cm 3cm,clip,width=1.0
    \textwidth]{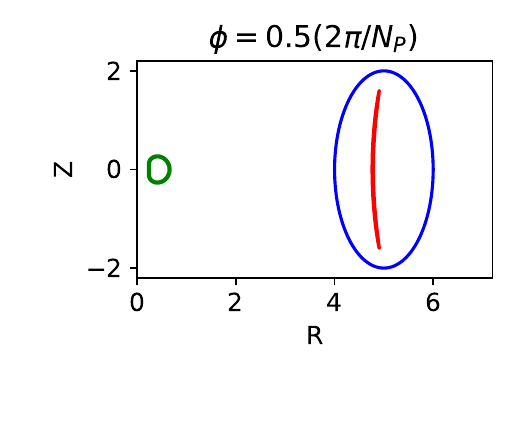}    
    \end{subfigure}
    \begin{subfigure}{0.62\textwidth}
    \includegraphics[trim=1.5cm 2cm 
    7.5cm 3cm,clip,width=0.49
    \textwidth]{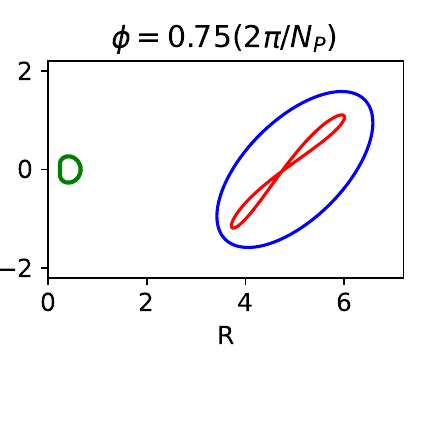}    
    \end{subfigure}
    \caption{Cross-sections of the initial (blue) and optimized surfaces obtained by minimizing the objective function given by \eqref{eq:volume_objective}. The surface obtained without the constraints ($\lambda_R = \lambda_{\mathcal{S}} = 0$) becomes self-intersecting (red), and the surfaced obtained with the constraints remains regular (green).}
    \label{fig:volume_opt_xc}
\end{figure}

We compare the surfaces obtained with and without ($\lambda_{\mathcal{S}} = \lambda_{R} = 0$) the constraint terms, shown in Figure \ref{fig:volume_opt_xc}. Without the constraint, the boundary begins to self-intersect. Although the volume is not well-defined for this surface, the discretized surface integral given by \eqref{eq:volume_surface_integral} is reduced from its initial value of 197.39 m$^3$ to 4.18 m$^3$ by minimizing the projection of the normal vector in the $\hat{\textbf{Z}}$ direction. With the constraint, the surface is reduced to an axisymmetric torus with a nearly circular cross-section centered at a major radius $R_0 = 0.43$ m with an averaged minor radius of $0.24$ m and a volume of 0.50 m$^3$. 

\begin{figure}
    \centering
    \begin{subfigure}[b]{0.49\textwidth}
    \centering
    \includegraphics[trim=2cm 2cm 1cm 3cm,clip,width=1.0
    \textwidth]{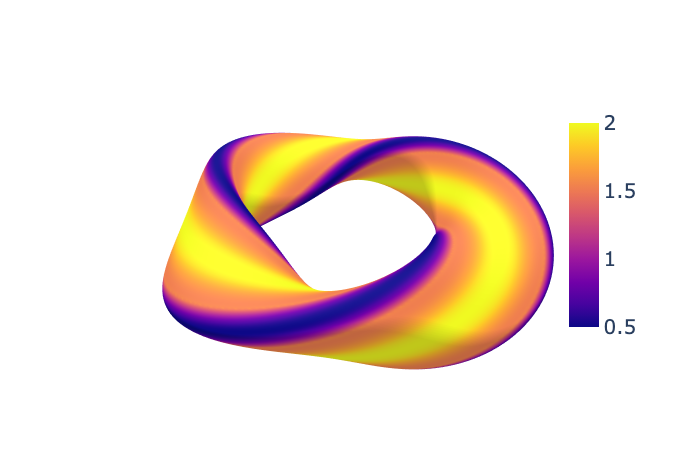}
    \caption{Initial $\rho_G$}
    \end{subfigure}
    \begin{subfigure}[b]{0.49\textwidth}
    \centering
    \includegraphics[trim=7cm 2.5cm 1cm 2cm,clip,width=1.0
    \textwidth]{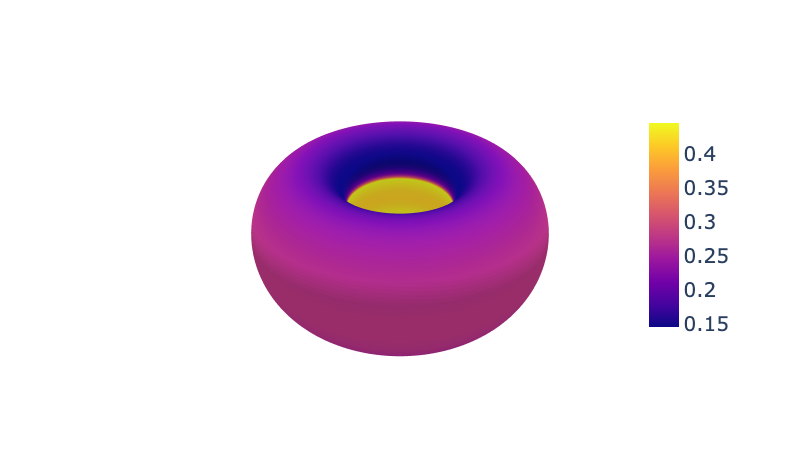}
    \caption{Optimized $\rho_G$}
    \end{subfigure}
    \caption{The global radius of curvature \eqref{eq:rho_G} of the (a) initial surface \eqref{eq:initial_volume} and (b) surface optimized for minimal volume with regularity constraints \eqref{eq:volume_objective}.}
    \label{fig:const_vol}
\end{figure}

In Figure \ref{fig:const_vol} we display the value of the global radius of curvature \eqref{eq:rho_G} for the initial and final surfaces. On the initial surface, $\rho_G$ is minimized at the endpoints of the semi-major axis where the global radius of curvature matches the local radius of curvature. The maximum value of $\rho_G$ is obtained in the region near the endpoint of the ellipse's semi-minor axis, where the point-tangent radius is minimized for $\textbf{x}_2$ on the opposite side of the semi-minor axis. The optimized boundary features decreased values of $\rho_G$ on the outboard side, where $\rho_G$ matches the local radius of curvature. We note that the minimum value of $\mathcal{S}_{\mathcal{C}}$ on the optimized slightly violates the constraint at 0.15 m due to the penalty formulation. 




\section{Optimization demonstrations}
\label{sec:demonstration}

\subsection{Target rotational transform}
\label{sec:iota_target}

We now discuss a demonstration of optimization to obtain a target rotational transform profile. We consider a target function which quantifies the difference between the rotational transform and a desired profile, $\iota_{\text{target}}(\psi)$,
\begin{align}
    f_{\iota} = \frac{1}{2} \int_{V_P} d\psi \, \left(\iota(\psi)-\iota_{\text{target}}(\psi) \right)^2,
    \label{eq:iota_target}
\end{align}
in order to obtain a low shear stellarator. The gradient of this objective function is obtained by computing an equilibrium with a perturbed toroidal current profile,
\begin{align}
\delta I_T(\psi) = w(\psi) (\iota(\psi) - \iota_{\text{target}}(\psi)),
\end{align}
as described in \citep{Antonsen2019}. We present a benchmark problem in Appendix \ref{app:target_iota} to demonstrate that the gradient-based optimizer can converge to a known minimum of an objective involving $f_{\iota}$ in a 2D space. 

While the VMEC code assumes nested surfaces such that the magnetic field is integrable, the solution to Laplace's equation with the prescribed boundary $S_P$ will not generally have continuously nested surfaces. We can, however, aim to obtain a vacuum field with a value of the rotational transform that improves the integrability of the ``real'' vacuum field. (Note that while we can run VMEC with prescribed $\iota(\psi)$ rather than $I_T(\psi)$, it will not generally be a vacuum field.)
We choose $\iota_{\text{target}}$ to be the ``most irrational'' noble between $p/q = 0/1$ and $p'/q'=1/2$ \citep{Greene1986},
\begin{align}
    \iota_{\text{target}} = \frac{p + \gamma p'}{q + \gamma q'} \approx 0.381966,
\end{align}
where $\gamma = (\sqrt{5} + 1)/2$ is the golden mean. This is a noble irrational, indicating that its path in the Farey tree is eventually alternating. At a given level in the Farey tree, the path that changes direction is said to be ``most irrational'' if a cantorus with this rotation number has the smallest flux of field line trajectories \citep{Meiss1992}. Thus the value of $\iota_{\text{target}}$ is chosen to have the smallest flux for rotation numbers between $p/q = 0/1$ and $p'/q' = 1/2$. According to the KAM theorem, invariant circles with sufficiently irrational frequencies persist under small perturbations. Thus choosing such a value for the rotational transform is likely to result in a large volume of magnetic surfaces. We remark that moderate shear may be desirable for certain stellarator design studies. Our adjoint formalism is quite flexible, and the choice of $\iota_{\text{target}}$ could be modified accordingly.

We define our objective function to be,
\begin{align}
    f(S_P) = f_{\iota} + \lambda_R f_R + \lambda_{\mathcal{S}} f_{\mathcal{S}},
\label{eq:iota_objective}
\end{align}
where $\lambda_{\mathcal{S}} = \lambda_{R} = 10^3$, $R_{\text{min}} = \mathcal{R} = 0.2$ m, and $w_R = w_{\mathcal{S}} = 10^{-2}$. We begin with a surface given by a rotating ellipse with a non-planar axis,
\begin{subequations}
\begin{align}
    R(\theta,\phi) = R_0 + R_{0,1} \cos(\phi) &+ 0.5a\left(\cos \left(\theta - N_{P} \phi\right) + \cos(\theta) \right) \nonumber \\
    &- 0.5 b \left(\cos\left(\theta - N_{P}\phi\right) -\cos(\theta)\right)
\end{align}
\begin{align}
    Z(\theta,\phi) = -R_{0,1} \sin(\phi)  &+ 0.5b\left(\sin(\theta) + \sin\left(\theta - N_{P}\phi\right) \right) \nonumber \\
    &+ 0.5 a \left(\sin(\theta) -\sin\left(\theta - N_{P} \phi\right)\right),
\end{align}
\label{eq:rotating_ellipse_torsion}
\end{subequations}
with $a = 2$, $b = 1$, $N_P = 3$, $R_0 = 5$, and $R_{0,1} = -0.5$. We consider a vacuum field with $p(\psi) = I_T(\psi) = 0$. 

\begin{figure}
    \centering
    \begin{subfigure}[b]{0.49\textwidth}
    \includegraphics[trim=0cm 0cm 0cm 0cm,clip,width=1.0\textwidth]{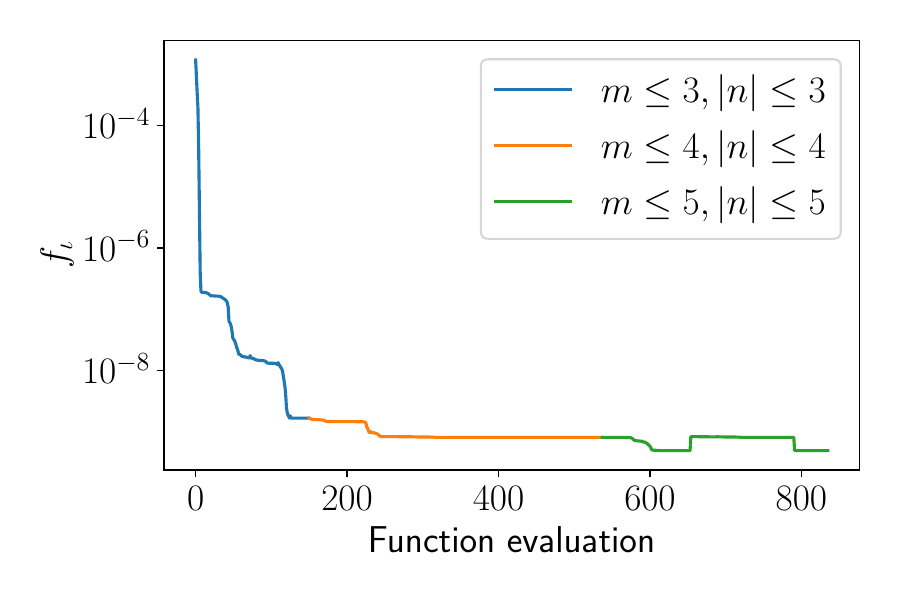}
    \caption{}
    \end{subfigure}    
    \begin{subfigure}[b]{0.49\textwidth}
    \includegraphics[trim=0cm 0cm 0cm 0cm,clip,width=1.0\textwidth]{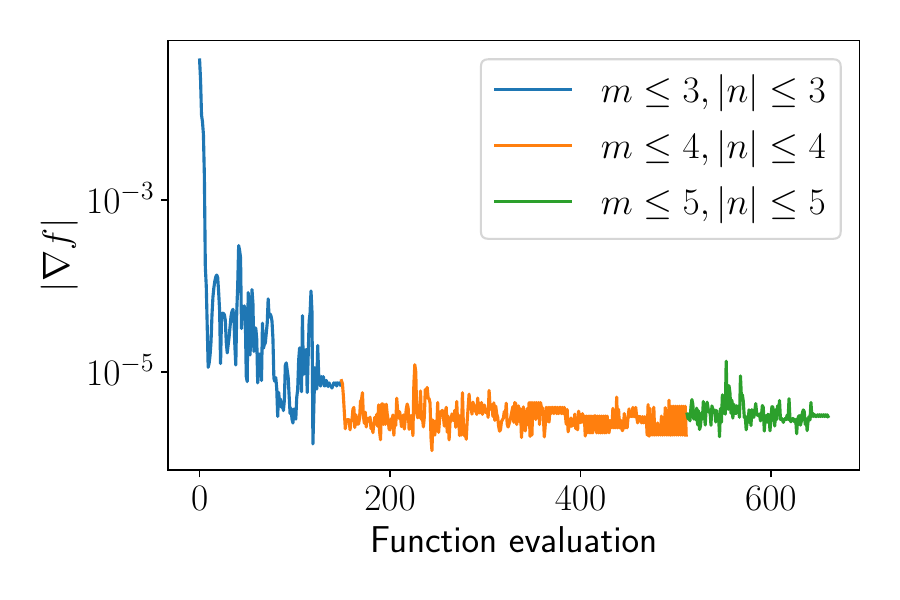}
    \caption{}
    \end{subfigure}
    \begin{subfigure}[b]{0.49\textwidth}
    \includegraphics[trim=0cm 0cm 0cm 0cm,clip,width=1.0\textwidth]{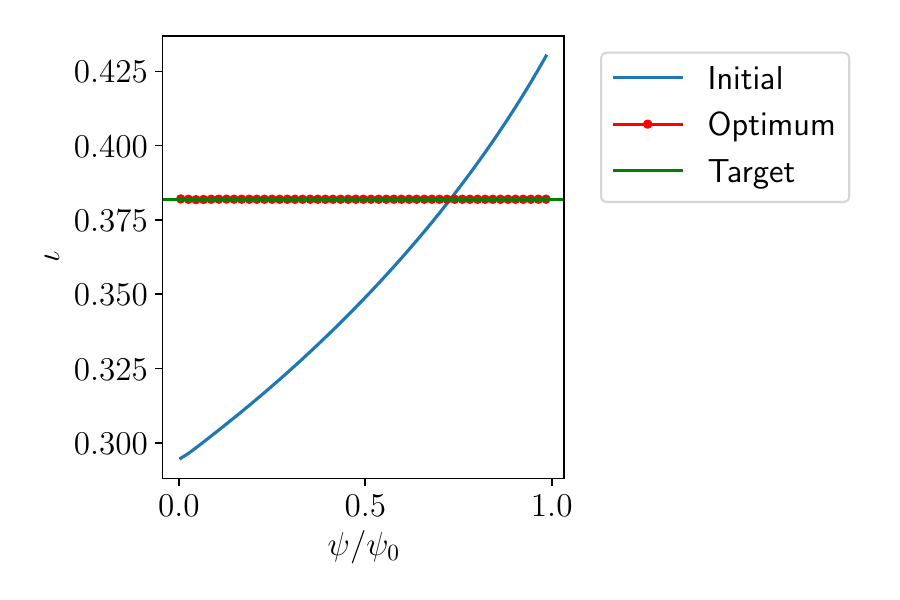}
    \caption{}
    \label{fig:iota_profile}
    \end{subfigure}
    \caption{The (a) $f_{\iota}$ objective value \eqref{eq:iota_target}, (b) L2 norm of the $f(S_P)$ objective \eqref{eq:iota_objective} gradient, and (c) the initial and final profiles of the rotational transform. In (a) and (b) the optimization was performed in a staged approach: first with respect to modes $m \le 3$ and $|n| \le 3$, then with respect to $m \le 4$ and $|n| \le 4$, and finally with respect to $m \le 5$ and $|n| \le 5$.}
    \label{fig:iota_convergence}
\end{figure}

To investigate the benefit of increasing the dimensionality of the optimization space, we optimize with respect to the boundary harmonics $m \le 3$, $|n| \le 3$. We then use this result to optimization with respect to $m \le 4$, $|n| \le 4$ and then with respect to $m \le 5$, $|n| \le 5$. In comparison with the result of the initial optimization in the low-dimensional space, we are able to reduce the objective function by 70\% (Figure \ref{fig:iota_convergence}). We obtain a rotational transform profile which very closely matches the target value with an objective value of $f_{\iota} = 4.93 \times 10^{-10}$. If the optimization space is further increased, the optimum is reduced by less than 1\%. For the further analysis of the optimization in this Section, we present results from the optimization with respect to the modes $m \le 5$ and $|n| \le 5$.

We note that we were not able to reduce the gradient norm to the requested tolerance of $10^{-8}$. This result can be attributed to approximations made in computing the gradient with the adjoint method. As discussed in \citep{Antonsen2019,Paul2020}, the adjoint equations are derived under the assumption that the VMEC code satisfies MHD force balance \eqref{eq:MHD}. However, there is always some residual error in force balance due to discretization error and the assumption of continuously nested flux surfaces. For these calculations, we converged to a force balance tolerance of $10^{-11}$ with 99 flux surfaces and mode numbers $|m| \le 11$, $|n| \le 11$. These resolution parameters were chosen to strike a balance between an accurate adjoint solution and efficiency of the optimization. Preliminary calculations indicate that increasing the resolution parameters over what was used in this work does not cause a significant change in the optimum. Furthermore, the linear adjoint solution is approximated by adding a small perturbation to the nonlinear MHD force balance. This technique is effectively a forward-difference approximation of the adjoint equation, which introduces an error that scales with the magnitude of the perturbation. When there are small errors in the gradient, the computed gradient may no longer be a descent direction, making convergence difficult near the optimum \citep{Dekeyser2014}. 

\begin{figure}
\centering
    \begin{subfigure}{0.49\textwidth}
    \includegraphics[trim=0cm 1cm 
    5cm 0cm,clip,width=0.70
    \textwidth]{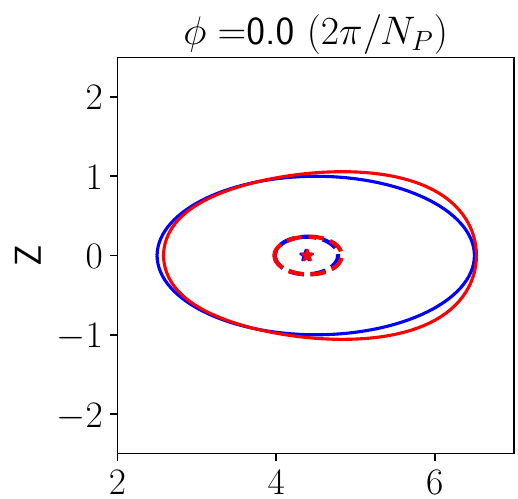}    
    \end{subfigure}
    \begin{subfigure}{0.49\textwidth}
    \includegraphics[trim=1cm 1cm 
    0cm 0cm,clip,width=1.0
    \textwidth]{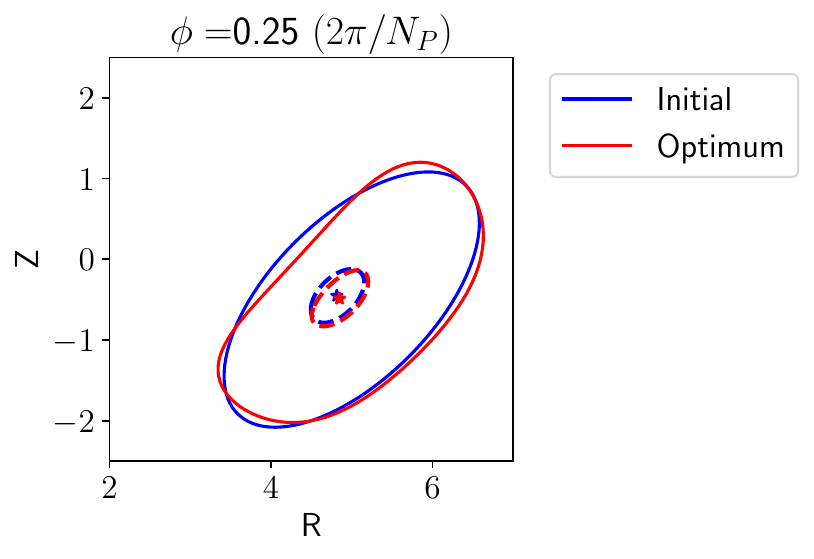}    
    \end{subfigure}
    \begin{subfigure}{0.49\textwidth}
    \includegraphics[trim=0cm 0cm 
    5cm 0cm,clip,width=0.70
    \textwidth]{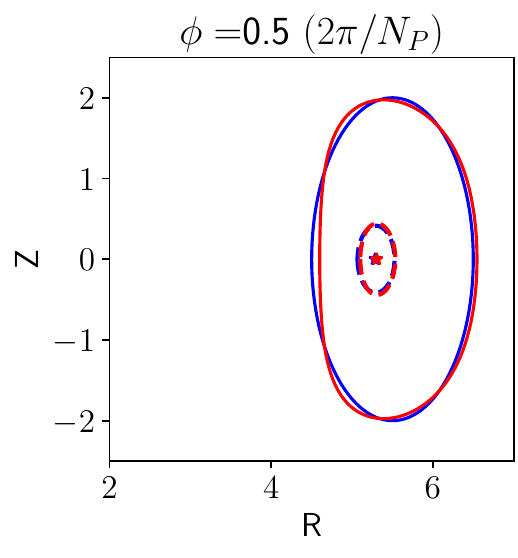}    
    \end{subfigure}
    \begin{subfigure}{0.49\textwidth}
    \includegraphics[trim=1cm 0cm 
    5cm 0cm,clip,width=0.61
    \textwidth]{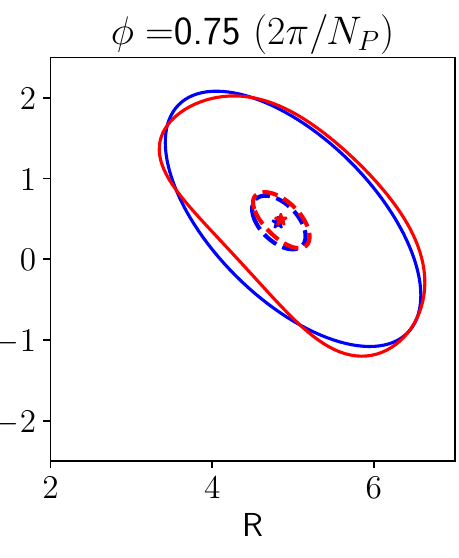}    
    \end{subfigure}
    \caption{Poloidal cross-sections of the initial boundary \eqref{eq:rotating_ellipse_torsion} and boundary which optimizes \eqref{eq:iota_objective} (solid) along with the surface at $\psi/\psi_0 = 0.06$ (dashed) and magnetic axis (star).}
    \label{fig:cross_section_iota}
\end{figure}

Nonetheless, we effectively eliminate the magnetic shear throughout the volume. In Figure \ref{fig:cross_section_iota} we display the initial and optimized boundaries (solid) along with an interior surface at $\psi/\psi_0 = 0.06$ (dashed) and the magnetic axis (star). We can consider some of the features of the optimized surface in consideration of the expression for the rotational transform near the magnetic axis \citep{Mercier1964,Helander2014}, which indicates that rotating ellipticity and torsion of the axis contribute to the on-axis transform in a vacuum field. We see that the interior surfaces become slightly more elliptical in order to increase $\iota$ near the axis, while the ellipticity of the boundary is slightly decreased, becoming more square. However, the torsion of the magnetic axis is maintained. 

\begin{figure}
    \centering
    \begin{subfigure}{0.49\textwidth}
    \includegraphics[trim=0cm 0cm 3cm 3cm,clip,width=1\textwidth]{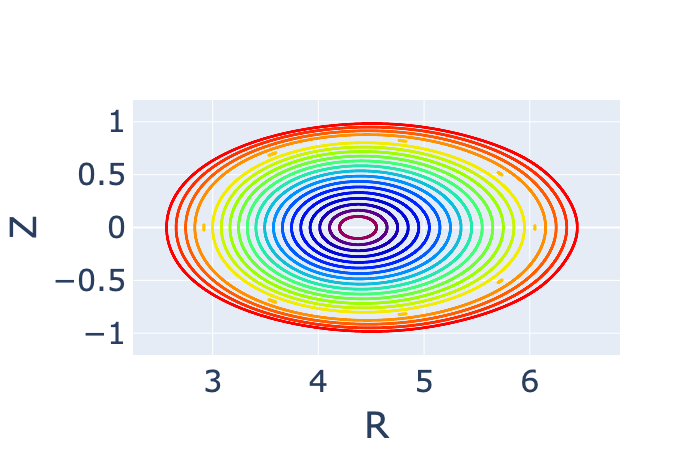}
    \caption{Initial}
    \label{fig:poincare_init}
    \end{subfigure}
    \begin{subfigure}{0.49\textwidth}
    \includegraphics[trim=4.5cm 0cm 3cm 3cm,clip,width=0.8\textwidth]{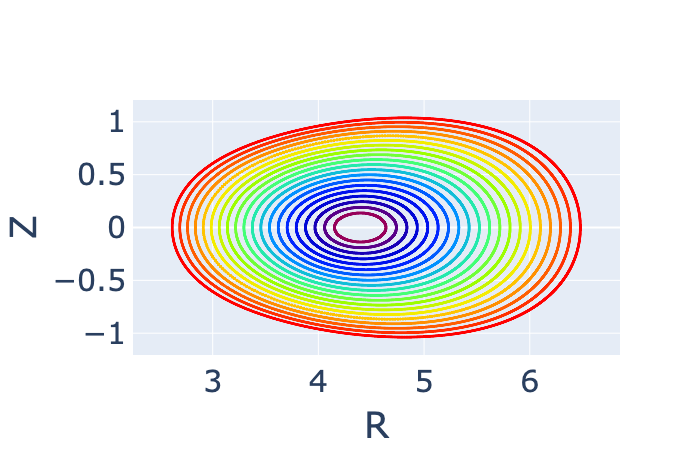}
    \caption{Optimized}
    \end{subfigure}
    \caption{Poincar\'{e} section computed from the SPEC vacuum field using the initial boundary \eqref{eq:rotating_ellipse_torsion} and boundary which minimizes \eqref{eq:iota_objective}.}
    \label{fig:poincare}
\end{figure}

To analyze the impact of low shear on the integrability of the field, we compute the vacuum field using the SPEC code \citep{Hudson2012}. The SPEC calculations are performed with a single volume with Beltrami parameter $\mu = 0$ such that the magnetic field satisfies $\nabla \times \textbf{B} = 0$ with a Neumann boundary condition \eqref{eq:neumann}. In Figure \ref{fig:poincare} we show a Poincar\'{e} section for the initial field, which has a small island chain at the $\iota = 3/8$ resonance. With the optimized boundary, we eliminate this island chain and obtain a large volume of nested surfaces.

\subsection{Magnetic well}
\label{sec:well}

We next consider an objective function which aims to obtain a magnetic well,
\begin{align}
f_w &=  \frac{\int_{V_P} d \psi \, \left(w_1(\psi) V'(\psi) - w_2(\psi) V'(\psi)\right)}{\int_{V_P} d \psi \, \left(w_1(\psi) V'(\psi) + w_2(\psi) V'(\psi)\right)},
\end{align}
with $w_1(\psi) = e^{-(\psi/\psi_0-1.0)^2/0.1^2}$ and $w_2(\psi) = e^{-(\psi/\psi_0)^2/0.1^2}$. When $V''(\psi) < 0$, a magnetic well is said to be present, which provides a stabilizing term in the Mercier criterion for interchange modes \citep{Mercier1974}. We can consider $f_w$ to be a normalized finite-difference approximation of $V''(\psi)$; thus minimization of $f_w$ is performed in order to achieve a magnetic well. A similar objective is employed in the ROSE code \citep{Drevlak2018}, computed from integration along a field line. The gradient of this objective function is obtained by computing an equilibrium with a perturbed pressure profile,
\begin{align}
    \delta p(\psi) &= \frac{w_1(\psi) - w_2(\psi) - f_w\left(w_1(\psi) + w_2(\psi)\right)}{\int_{V_P} d^3 x \, \left(w_1(\psi) +  w_2(\psi)\right)},
\end{align}
as described in \citep{Paul2020}.

We define our objective function to be,
\begin{align}
    f(S_P) = f_w + \frac{\lambda_V}{2}\left(V_P - V_P^{\text{init}}\right)^2  +  \lambda_{\mathcal{S}} f_{\mathcal{S}} + \lambda_R f_R
    + \lambda_{\kappa}  f_{\kappa} + \lambda_{\overline{\kappa}} f_{\overline{\kappa}},
    \label{eq:well_objective_function}
\end{align}
with $\lambda_{\mathcal{S}} = 10^{-3}$, $\lambda_{R} = 10^{2}$, $w_{\mathcal{S}} = w_{R} = 10^{-2}$, $\kappa_{\max} \equiv \kappa_{\max,1} = \kappa_{\max,2} = 7$ m$^{-1}$, $w_{\kappa} = 1$, $\lambda_{\overline{\kappa}}= \lambda_V = 1$, and $\lambda_{\kappa} = 10^{2}$. Here $V_{P}^{\text{init}}$ is the volume of the initial surface. To determine the importance of the maximum curvature regularization term, we optimize with four sets of parameters as described in Table \ref{tab:well}. The values $\lambda_{\kappa} = 10^2$ and $\kappa_{\max} = 7$ m$^{-1}$ were chosen to balance the curvature and well metrics.

\begin{table}
    \centering
    \begin{tabular}{ c c c c }
       $\lambda_{\kappa}$ & $\kappa_{\max}$ & $\max(|\kappa_1|,|\kappa_2|)$ & $f_w$ \\ \hline
       $10^2$ & 7 m$^{-1}$ & 6.55 m$^{-1}$ & -0.21 \\
       $10^3$ & 7 m$^{-1}$ & 6.36 m$^{-1}$ & -0.07 \\
       $10^0$ & 8 m$^{-1}$ & 7.84 m$^{-1}$ &  -0.19 \\
       $10^0$ & 10 m$^{-1}$ & 9.95 m$^{-1}$ & -0.16 
    \end{tabular}
    \caption{Parameters used for the optimization of the magnetic well and the resulting values of the optimum objective function \eqref{eq:well_objective_function}.}
    \label{tab:well}
\end{table}

In addition to the regularization terms, we include a term in the objective function which penalizes a change in the volume, as an increase in the inverse aspect ratio can yield a Shafranov shift (\S 3.7 in \citep{Wesson2011}). This shift in the flux surfaces toward a smaller major radius implies that the volume of a flux surface increases less rapidly than its cross-sectional area. Furthermore, the flux through a surface increases more rapidly than its cross-sectional area because the geometric center is moving into a region of increased field strength (assuming the field is mostly toroidal). Thus the volume increases less rapidly than the flux, and a negative value of $V''(\psi)$ can be achieved \citep{Taylor1965}. 

\begin{figure}
\centering
    \begin{subfigure}{0.49\textwidth}
    \includegraphics[trim=1.5cm 1.2cm 
    4cm 0cm,clip,width=1.0
    \textwidth]{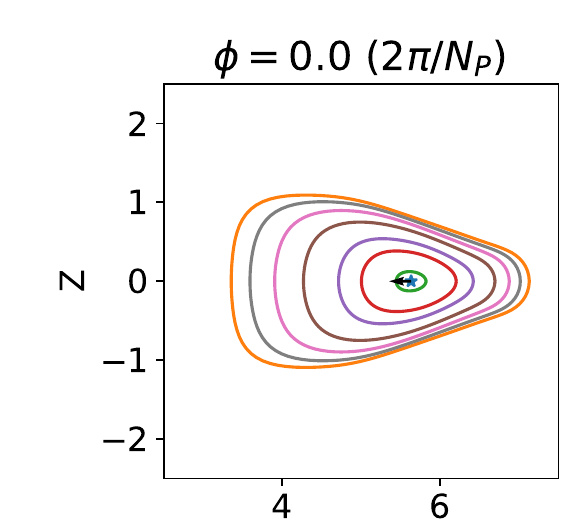}    
    \end{subfigure}
    \begin{subfigure}{0.49\textwidth}
    \includegraphics[trim=4cm 1.2cm 
    4cm 0cm,clip,width=0.74
    \textwidth]{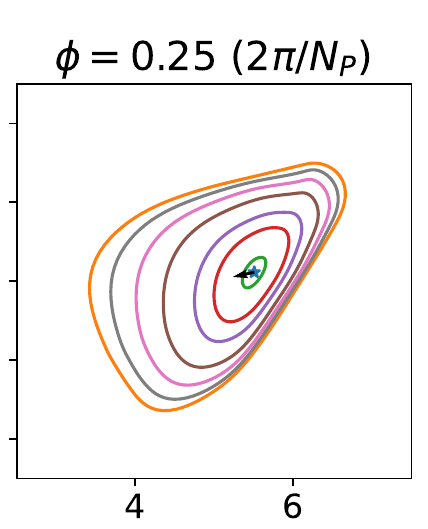}    
    \end{subfigure}
    \begin{subfigure}{0.49\textwidth}
    \includegraphics[trim=1.5cm 0cm 
    4cm 0cm,clip,width=1.0
    \textwidth]{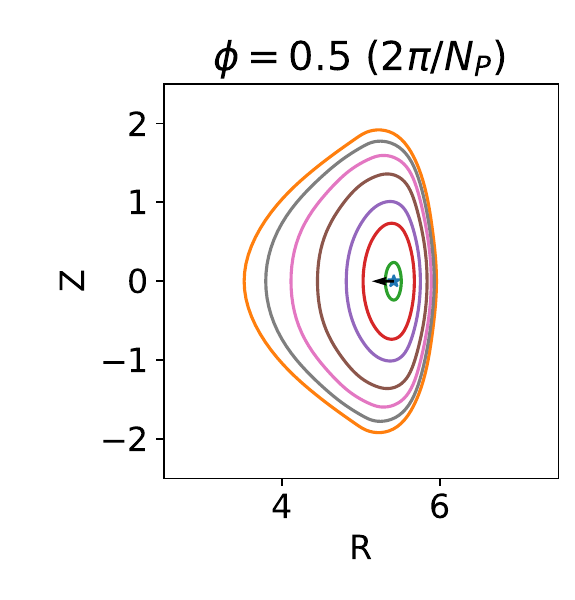}    
    \end{subfigure}
    \begin{subfigure}{0.49\textwidth}
    \includegraphics[trim=4cm 0cm 
    4cm 0cm,clip,width=0.74
    \textwidth]{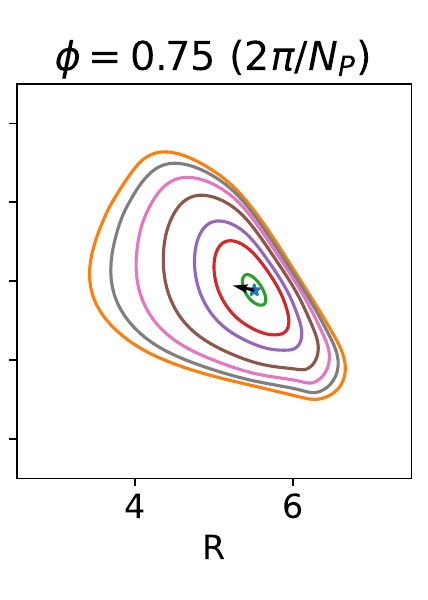}    
    \end{subfigure}
    \caption{Cross sections of the boundary (orange) optimized for the magnetic well \eqref{eq:well_objective_function} with regularization on the curvature $(\lambda_{\kappa} = 10^2,  \lambda_{\overline{\kappa}} = 1)$. Several interior surfaces are shown, along with the location of the magnetic axis (blue star) and axis normal vector (black arrow). The optimized boundary features rotating triangularity and elongation.}
    \label{fig:cross_section_well}
\end{figure}

We begin with a boundary given by a rotating ellipse \eqref{eq:rotating_ellipse} with $a = 2$, $b = 1$, $R_0 = 5$, and $R_{0,1} = 0$. We optimize the boundary with respect to the modes $m \le 10$ and $|n| \le 10$. The equilibrium is computed with $p(\psi) = 0$ and $I_T(\psi) = 0$ such that a vacuum field is considered. We arrive at the boundary given in Figure \ref{fig:cross_section_well}. With the addition of the volume constraint, the aspect ratio remains roughly constant (3.54 for the initial boundary and 3.46 for the optimized boundary), so that the well is provided by the shaping of the boundary rather than the Shafranov shift that arises due to the inverse aspect ratio.  We note that the optimized surface features triangularity that rotates from outward-pointing with horizontal elongation to inward-pointing with vertical elongation.

\begin{figure}
\centering
    \begin{subfigure}{0.49\textwidth}
    \includegraphics[trim=1.5cm 1.2cm 
    4cm 0cm,clip,width=1.0
    \textwidth]{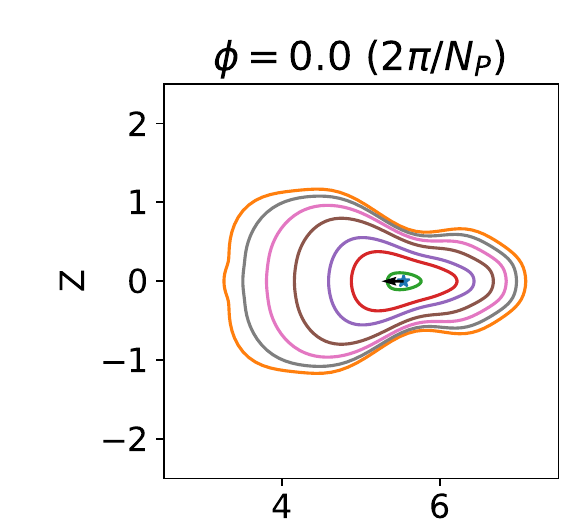}    
    \end{subfigure}
    \begin{subfigure}{0.49\textwidth}
    \includegraphics[trim=4cm 1.2cm 
    4cm 0cm,clip,width=0.74
    \textwidth]{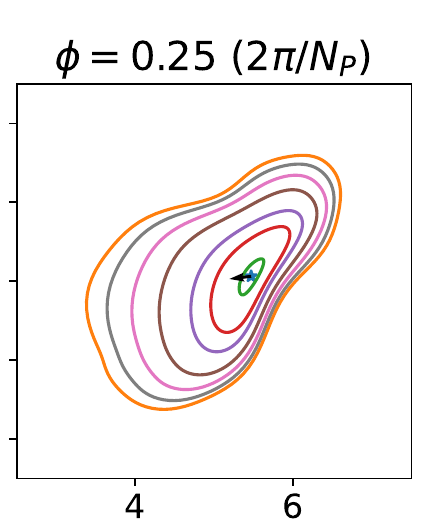}    
    \end{subfigure}
    \begin{subfigure}{0.49\textwidth}
    \includegraphics[trim=1.5cm 0cm 
    4cm 0cm,clip,width=1.0
    \textwidth]{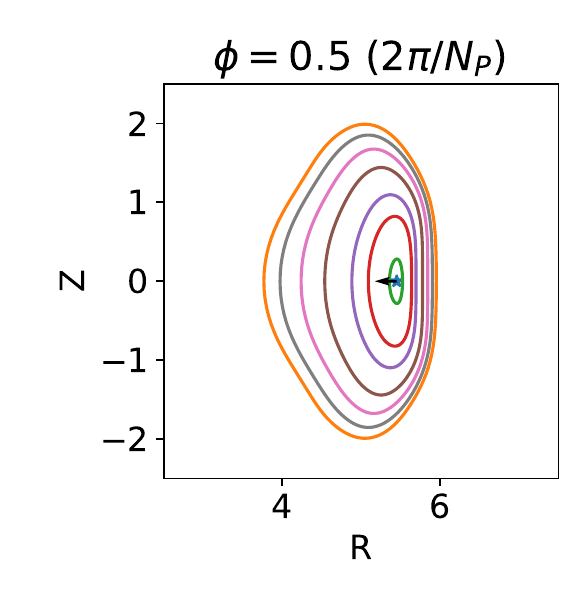}    
    \end{subfigure}
    \begin{subfigure}{0.49\textwidth}
    \includegraphics[trim=4cm 0cm 
    4cm 0cm,clip,width=0.74
    \textwidth]{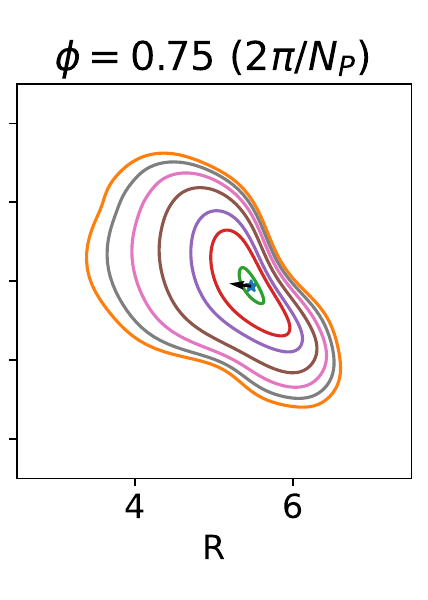}    
    \end{subfigure}
    \caption{Cross sections of the boundary optimized for the magnetic well \eqref{eq:well_objective_function} without regularization on the principal curvature $(\lambda_{\kappa} = \lambda_{\overline{\kappa}} = 0)$. Several interior surfaces are shown, along with the location of the magnetic axis (blue star) and axis normal vector (black arrow). The optimized boundary features a ``dimple'' feature at the $\phi = 0$ plane and regions of increased concavity.}
    \label{fig:cross_section_well_unconstrained}
\end{figure}

To demonstrate the effect of the curvature regularization terms, we examine the optimization with $\lambda_{\kappa} = \lambda_{\overline{\kappa}} = 0$.  We converge to the boundary shown in Figure \ref{fig:cross_section_well_unconstrained}. As can be seen, regions of large curvature are exhibited, including a dimple-like feature on the inboard side and a concave ``pinching'' feature. Again the aspect ratio remains roughly constant (3.46).

We can understand the geometric dependence of the magnetic well by considering the expression for $V''(\psi)$ on the axis that arises from the near-axis expansion in the inverse coordinate representation \citep{Landreman2020b},
\begin{align}
    V''(\psi) \propto \int_0^{2\pi} d \varphi_B \, \frac{1}{B_0^4}\left[3 (B_{1s}^2 + B_{1c}^2) - 4 B_0 B_{20}\right],
\end{align}
where $\varphi_B$ is the Boozer toroidal angle and we have made the assumption of a vacuum field. The field strength near the axis is expanded as,
\begin{multline}
    B = B_0(\varphi_B) + r \left[ B_{1c}(\varphi_B) \cos(\vartheta_B) + B_{1s}(\varphi_B) \sin(\vartheta_B)\right] \\ + r^2 \left[B_{20}(\varphi_B) + B_{2c} \cos(2 \vartheta_B) + B_{2s} \sin(2 \vartheta_B) \right] + \mathcal{O}\left(r^3\right),
\end{multline}
where $r \propto \sqrt{\psi}$ is the effective minor radius and $\vartheta_B$ is the Boozer poloidal angle. 

As can be seen from (A34) in \citep{Landreman2019}, the poloidally-independent shift in the flux surface in the normal direction, $X_{20}$, increases proportional to $B_{20}$. Here, the normal vector is $\hat{\textbf{n}} = \textbf{r}''(l)/|\textbf{r}''(l)|$, where $\textbf{r}(l)$ is the position vector along the magnetic axis parameterized by the length. In the axisymmetric limit, a positive value of $X_{20}$ indicates a net shift of the surfaces' geometric center toward a smaller major radius. This correlation between $X_{20}$ and $B_{20}$ is consistent with the statement that a Shafranov shift provides a magnetic well. In Figures \ref{fig:cross_section_well} and \ref{fig:cross_section_well_unconstrained} we display the magnetic axis (blue star) and normal vector (black arrow) along with the shapes of several magnetic surfaces for the optimized configurations. While both the unconstrained and constrained optima feature a net shift in the geometric center of the flux surfaces in the normal direction, this is achieved with vastly different shaping of the boundary. Interestingly, both configurations feature negative triangularity at the $\phi = 0.5$ $(2\pi/N_P)$ plane. 

We note that $B_{20}$ arises due to other shaping components of the surface \citep{Landreman2019}. In axisymmetry, negative values of $B_{2c}$ contribute to positive (outward-pointing) triangularity ($X_{2c}<0$) ((B11) in \citep{Landreman2020a}). Assuming stellarator symmetry, negative values of $B_{2c}$ contribute to positive values of $B_{20}$ if the surface is vertically elongated, and positive values of $B_{2c}$ contribute to positive values of $B_{20}$ if the surface is horizontally elongated. This implies that positive triangularity coupled with vertical elongation or negative triangularity coupled with horizontal elongation contributes to the magnetic well in axisymmetry. These trends are consistent with the stability analysis of oblate plasmas with negative triangularity \citep{Pogutse1982,Kesner1995,Medvedev2015}. In 3D, the connection between $B_{20}$ and $B_{2c}$ is more complicated ((A41)-(A42) in \citep{Landreman2019}). Nonetheless, we note that the magnetic well near the axis arises at second order in the expansion parameter, which includes the effects of ellipticity and triangularity. Thus it is not surprising that the optimized boundary with curvature constraints features rotating ellipticity and triangularity.

\begin{figure}
    \centering
    \begin{subfigure}[b]{0.49\textwidth}
    \includegraphics[trim=0cm 0cm 0cm 0cm,clip,width=1.0\textwidth]{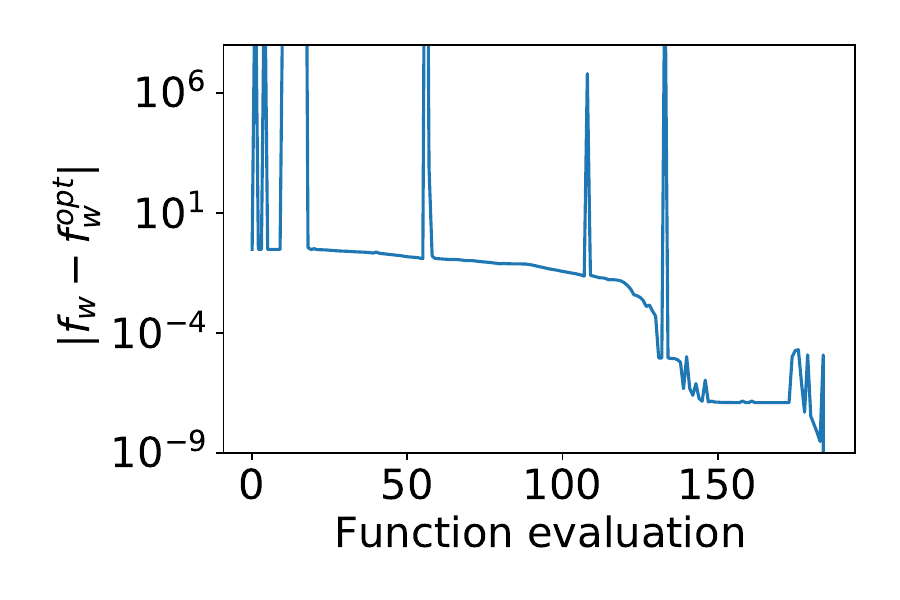}
    \caption{Constrained}
    \end{subfigure}    
    \begin{subfigure}[b]{0.49\textwidth}
    \includegraphics[trim=0cm 0cm 0cm 0cm,clip,width=1.0\textwidth]{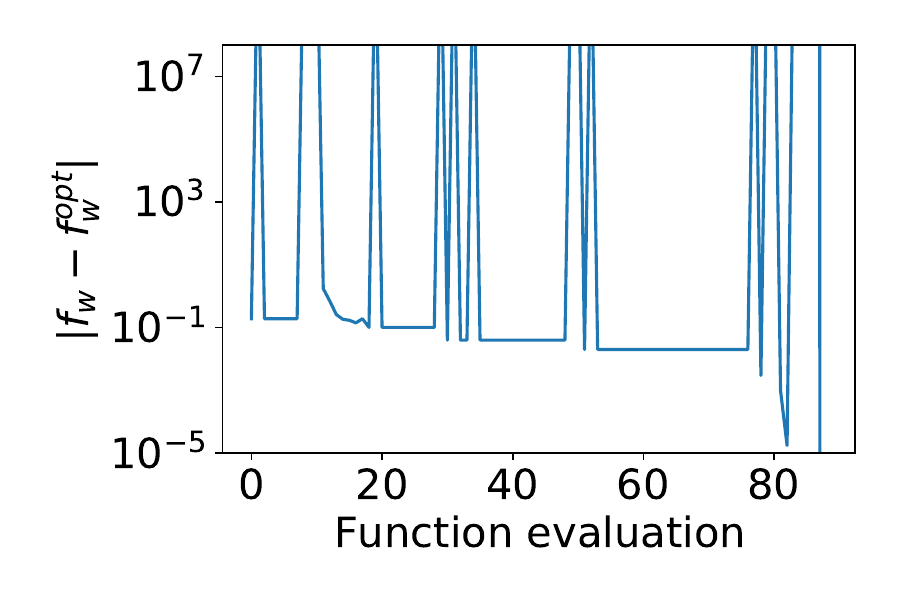}
    \caption{Unconstrained}
    \end{subfigure}    
    \begin{subfigure}[b]{0.49\textwidth}
    \includegraphics[trim=0cm 0cm 0cm 0cm,clip,width=1.0\textwidth]{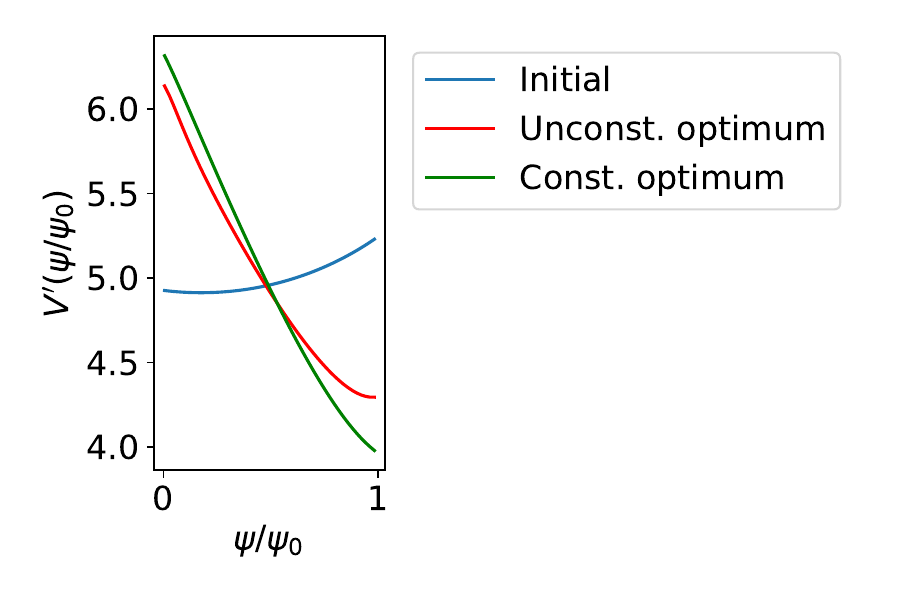}
    \caption{}
    \end{subfigure}
    \caption{Convergence of the optimization of the objective function \eqref{eq:well_objective_function} (a) with the inclusion of the curvature constraints ($\lambda_{\kappa} = 10^2, \lambda_{\overline{\kappa}} = 1$) and (b) without the curvature constraints ($\lambda_{\kappa} = \lambda_{\overline{\kappa}} = 0$). With the inclusion of the curvature constraint, we achieve a deeper magnetic well throughout the volume (c).}
    \label{fig:well_convergence}
\end{figure}

In comparing the convergence of the optimization with and without the curvature constraints (Figure \ref{fig:well_convergence}), we see that the presence of the curvature objective prevents some failures of the VMEC code. Each function evaluation that resulted in a VMEC failure is visualized as a ``spike'' that extends above $|f_w - f^{\text{opt}}_w| = 1$, as we artificially set the value of the objective function to $10^{12}$ at these points. While the constrained optimization still features some VMEC failures (12 vs. 19), the inclusion of the curvature constraints improves the convergence toward the optimum, and a deeper magnetic well is achieved throughout the volume. 


\subsection{Quasisymmetry}
\label{sec:ripple}

We next consider optimization for quasisymmety near the magnetic axis. This is quantified through the objective,
\begin{align}
    f_{QS} = \frac{\int_{V_P} d^3 x \, w(\psi) (B - \overline{B})^2}{\int_{V_P} d^3 x \, B^2},
    \label{eq:axis_ripple}
\end{align}
where $\overline{B} = \int_{V_P} d^3 x \, w(\psi) B$ and we take $w(\psi) = e^{-(\psi/\psi_0)^2/0.1^2}$. This objective aims to make the field strength constant on the magnetic axis, which is a feature of quasi-axisymmetric and quasi-helically symmetric equilibria. A similar objective function has also been included in optimization for energetic particle confinement \citep{Drevlak2014}. The gradient of this objective function is obtained by computing an equilibrium with the addition of an anisotropic pressure tensor, $\nabla \cdot \textbf{P} = p_{||} \hat{\textbf{b}}\hat{\textbf{b}} + p_{\perp} \left(\textbf{I} -\hat{\textbf{b}}\hat{\textbf{b}}\right)$ with,
\begin{subequations}
\begin{align}
    p_{||}(\psi,B) &= \frac{w(\psi) (B - \overline{B})^2 - f_{QS}w(\psi) B^2}{\int_{V_P} d^3 x \, w(\psi) B^2} \\
    p_{\perp} &= p_{||} - B \partder{p_{||}(\psi,B)}{B},
\end{align}
\end{subequations}
as described in \citep{Paul2020}.

We take our objective function to be,
\begin{align}
    f(S_P) = f_{QS} + \lambda_{\iota} f_{\iota} + \lambda_{\overline{\kappa}} f_{\overline{\kappa}} + \lambda_{\kappa} f_{\kappa} + \lambda_{\mathcal{S}} f_{\mathcal{S}} + \lambda_R f_R,
    \label{eq:ripple_objective}
\end{align}
where $\iota_{\text{target}}(\psi)$ is taken to be the initial rotational transform profile. We also take $\lambda_{\iota} = 1$, $\lambda_{\overline{\kappa}} = 10^{-3}$, $\lambda_{\kappa} = 5 \times 10^{-2}$, $\kappa_{\text{max},1} = \kappa_{\text{max},2} = 10$ m$^{-1}$, $w_{\kappa} = 1$, $w_{\mathcal{S}} = w_R = 10^{-1}$, $\lambda_R = 10^3$, $\lambda_{\mathcal{S}} = 10^{-1}$, $\mathcal{R} = 0.5$ m, and $R_{\text{min}} = 0.2$ m. The additional constraint on the rotational transform is required to prevent the surface from becoming axisymmetric to reduce the toroidal ripple on the axis. 

\begin{figure}
    \centering
    \begin{subfigure}[b]{0.49\textwidth}
    \includegraphics[width=0.8\textwidth]{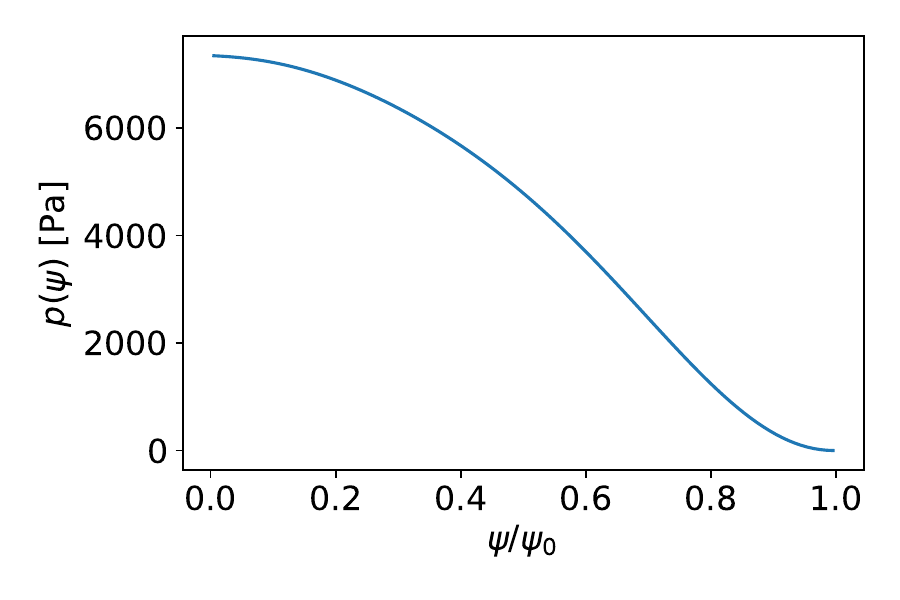}
    \caption{}
    \end{subfigure}
    \begin{subfigure}[b]{0.49\textwidth}
    \includegraphics[width=0.8\textwidth]{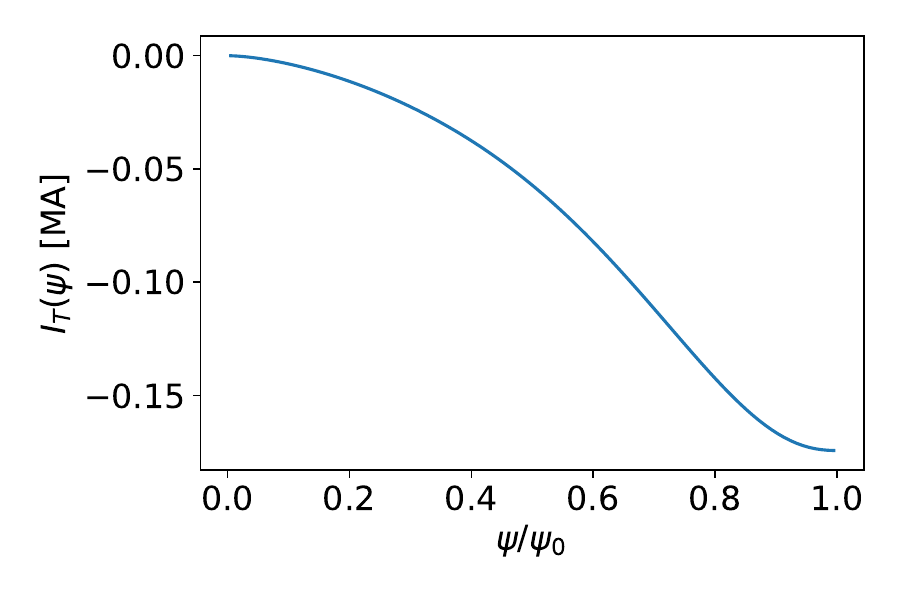}
    \caption{}
    \end{subfigure}
    \caption{The (a) pressure and (b) integrated toroidal current profiles used for the optimization of quasisymmetry on the axis \eqref{eq:ripple_objective}.}
    \label{fig:profiles}
\end{figure}

We begin with a rotating ellipse boundary with torsion, given by,
\begin{subequations}
\begin{align}
    R(\theta,\phi) = R_0 + R_{0,1} \cos(\phi) + R_{0,2}\cos(2\phi) &+ 0.5a\left(\cos \left(\theta - N_{P} \phi\right) + \cos(\theta) \right) \nonumber \\
    &- 0.5 b \left(\cos\left(\theta - N_{P}\phi\right) -\cos(\theta)\right)
\end{align}
\begin{align}
    Z(\theta,\phi) = R_{0,1} \sin(\phi) + Z_{0,2}\cos(2\phi)  &+ 0.5b\left(\sin(\theta) + \sin\left(\theta - N_{P}\phi\right) \right) \nonumber \\
    &+ 0.5 a \left(\sin(\theta) -\sin\left(\theta - N_{P} \phi\right)\right),
\end{align}
\label{eq:initial_ripple}
\end{subequations}
with $R_0 = 5$, $R_{0,1} = R_{0,2} = 0.3$, $a = 2$,  $b = 1$, and $N_P = 3$. We perform optimization at $\beta = 1.2$ \% with the profiles shown in Figure \ref{fig:profiles}. We optimize with respect to the modes $m \le 10$ and $|n| \le 10$ in a staged approach: first optimizing with respect to $m \le 3$ and $|n| \le 3$, then with respect to $m \le 6$ and $|n| \le 6$, then with respect to the full set of modes. 

\begin{figure}
    \centering
    \begin{subfigure}[b]{0.49\textwidth}
    \includegraphics[width=0.8\textwidth]{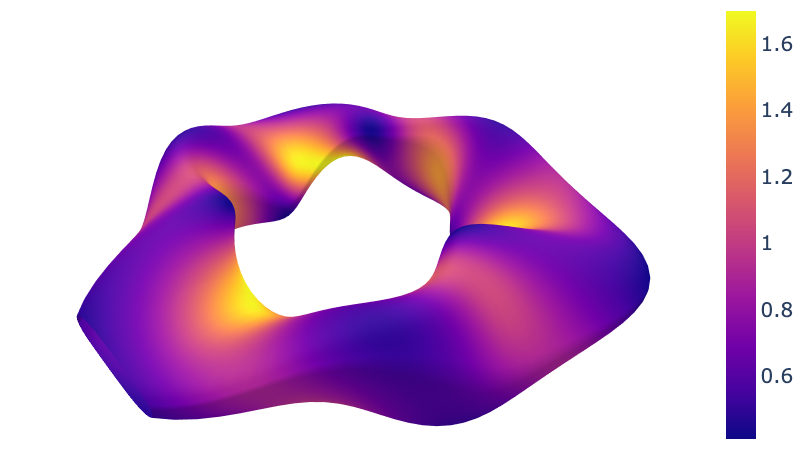}
    \caption{Initial}
    \end{subfigure}
    \begin{subfigure}[b]{0.49\textwidth}
    \includegraphics[width=0.8\textwidth]{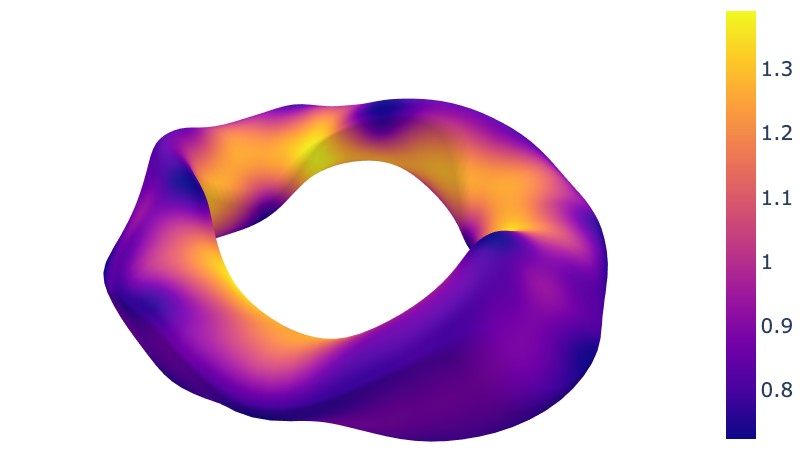}
    \caption{Optimum}
    \end{subfigure}
    \caption{The magnetic field strength on the (a) initial boundary \eqref{eq:initial_ripple} and (b) boundary optimized for quasisymmetry on the axis \eqref{eq:ripple_objective}.}
    \label{fig:ripple_B}
\end{figure}

We display the magnetic field strength on the initial and optimized boundaries in Figure \ref{fig:ripple_B}. We note the initial configuration features a large toroidal variation of the field strength, with increased field strength near the ``corners'' that arise due to the axis's large torsion. Although the axis ripple figure of merit \eqref{eq:axis_ripple} reduces the toroidal variation of the field strength on the axis, which must vanish in both quasi-axisymmetry and quasi-helical symmetry, we find that the optimized magnetic field is driven closer to quasi-axisymmetry.

\begin{figure}
    \centering
    \begin{subfigure}[b]{0.26\textwidth}
    \includegraphics[trim=0cm 0cm 6cm 0cm ,clip,width=1.0\textwidth]{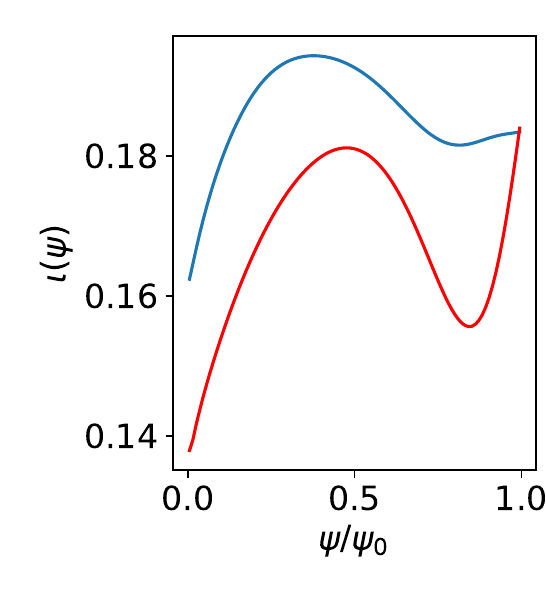}
    \caption{}
    \end{subfigure}
    \begin{subfigure}[b]{0.26\textwidth}
    \includegraphics[trim=0cm 0cm 6cm 0cm ,clip,width=1.0\textwidth]{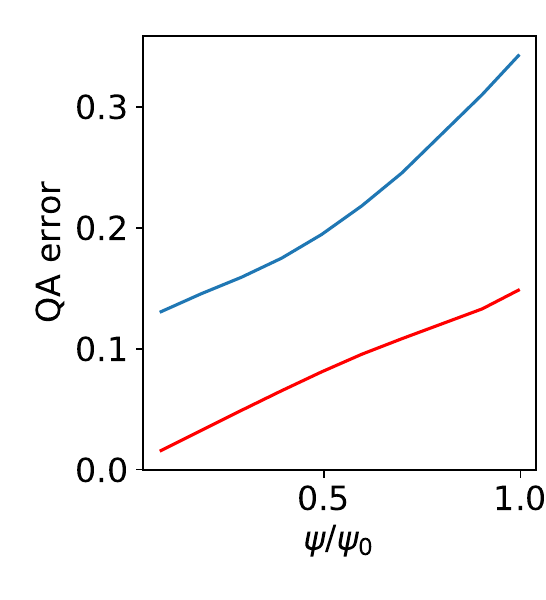}
    \caption{}
    \end{subfigure}
    \begin{subfigure}[b]{0.443\textwidth}
    \includegraphics[width=1.0\textwidth]{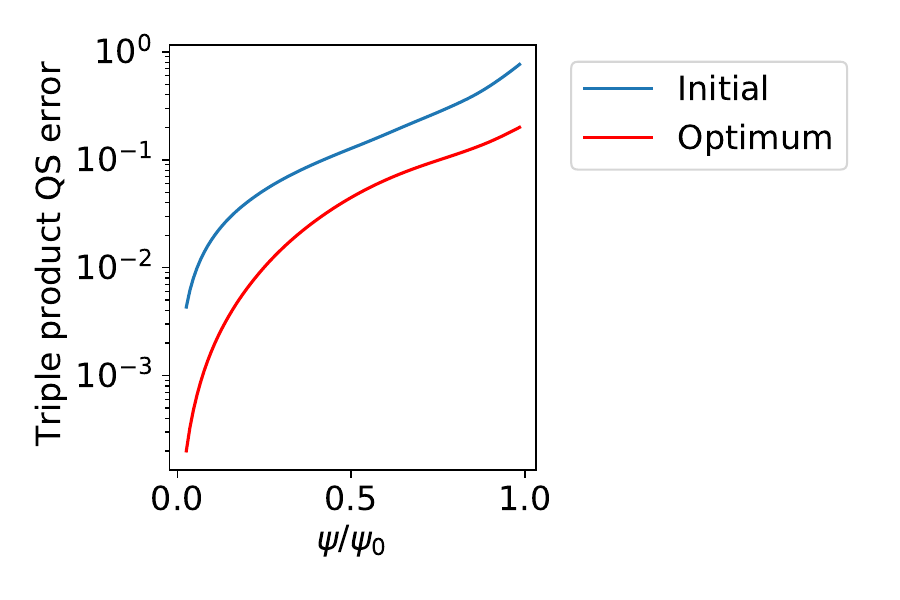}
    \caption{}
    \end{subfigure}
    \caption{The (a) rotational transform, (b) QA error \eqref{eq:qa_error}, and (c) triple product error \eqref{eq:triple_product} for the initial boundary \eqref{eq:initial_ripple} and boundary optimized for small magnetic ripple on the axis \eqref{eq:ripple_objective}.}
    \label{fig:ripple_opt}
\end{figure}

We remark that a design based solely on the objective \eqref{eq:axis_ripple} may be limited in that quasisymmetry is not targeted away from the axis. We therefore quantify the departure from quasisymmetry in the two configurations in two additional ways. The first is through the figure of merit,
\begin{align}
    f_{TP}(\psi) &=\frac{\left \langle \left \rvert \partder{B}{\theta} \partder{}{\phi} \left(\textbf{B} \cdot \nabla B \right) - \partder{B}{\phi} \partder{}{\theta}\left(\textbf{B} \cdot \nabla B\right) \right \rvert \right \rangle_{\psi}}{\langle B^3 \rangle_{\psi}/r_{\text{minor}}(\psi)},
    \label{eq:triple_product}
\end{align}
where $r_{\text{minor}}(\psi) = \sqrt{A(\psi)/\pi}$ is the averaged minor radius of a flux surface where $A(\psi)$ is the toroidally averaged cross-sectional area. (This is the same definition of the minor radius used in the VMEC code.) The flux-surface average of a quantity $Q$ is,
\begin{align}
\langle Q \rangle_{\psi} = \frac{\int_0^{2\pi} d\theta \int_0^{2\pi} d \phi \, Q \sqrt{g}}{\int_0^{2\pi} d \theta \int_0^{2\pi} d \phi \, \sqrt{g}},
\end{align}
where $\sqrt{g} = \left(\nabla \psi \times \nabla \theta \cdot \nabla \phi\right)^{-1}$ is the flux coordinate Jacobian. Eq. \eqref{eq:triple_product} is a normalized figure of merit which employs the triple product form for quasisymmetry \citep{Helander2014,Rodriguez2020},
\begin{align}
    \nabla \psi \times \nabla B \cdot \nabla \left(\textbf{B} \cdot \nabla B \right) = 0,
\end{align}
and allows us to quantify the quasisymmetry error without specifying the helicity of the symmetry.  We also perform a Boozer coordinate ($\psi$,$\vartheta_B$,$\varphi_B$) transformation \citep{Sanchez2000} to obtain the Fourier harmonics of the field strength,
\begin{align}
    B(\psi,\vartheta_B,\varphi_B) = \sum_{m,n} B_{m,n} \cos(m \vartheta_B - n \varphi_B).
\end{align}
The deviation from quasi-axisymmetry can then be quantified as,
\begin{align}
    f_{QA}(\psi) &= \sqrt{\frac{\sum_{m,n\ne 0} B_{m,n}(\psi)^2}{\sum_{m,n}B_{m,n}(\psi)^2}}.
    \label{eq:qa_error}
\end{align}
Although quasisymmetry was only targeted on the axis,  we reduce the quasisymmetry error throughout the volume with respect to both metrics without introducing a reduction in the rotational transform (Figure \ref{fig:ripple_opt}).


\section{Conclusions}
\label{sec:conclusions}

In this work, we have provided the first example of gradient-based, fixed-boundary optimization of stellarator equilibria. We provide examples of several equilibria obtained with the ALPOpt tool, including a vacuum field with ultra-low magnetic shear. Furthermore, we have identified regularization terms for fixed-boundary optimization that prevent self-intersection and reduce the surface curvature. These regularization terms improve the convergence toward the optimum and may reduce the required coil complexity. Another approach to reducing coil complexity would be to incorporate metrics related to the properties of the current potential solution on a uniformly offset winding surface \citep{Carlton2020}. 

The availability of derivative information enables the optimization in high-dimensional spaces. We present the optimization of the boundary with respect to the set of boundary harmonics $m \le 10$, $|n| \le 10$ (441 modes). This dimensionality is significantly larger than that of previous optimized efforts. For example, ESTELL was optimized with respect to the modes $m \le 4$, $|n| \le 4$ \citep{Drevlak2013} and NCSX was optimized with respect to $m \le 6$ and $|n| \le 4$ \citep{Zarnstorff2001}. Such an increase in the optimization space may enable further refinement of an optimum, as demonstrated in \S \ref{sec:iota_target}.

Gradient information of equilibrium quantities is obtained using an adjoint method \citep{Antonsen2019,Paul2020} which requires solving a modified equilibrium problem. For the objectives presented in this work -- the rotational transform (\S \ref{sec:iota_target}), the magnetic well (\S \ref{sec:well}), and near-axis quasisymmetry (\S \ref{sec:ripple}) -- the adjoint equilibrium problem requires the addition of a perturbation to the toroidal current profile, pressure profile, and the addition of an anisotropic pressure tensor of a specific form. For other figures of merit, the adjoint equilibrium problem requires the addition of a bulk force of a different form, such as an anisotropic pressure that cannot be handled by the variational principle employed by the ANIMEC code \citep{Cooper19923d}. In this case, rather than approximating the linearized adjoint problem by adding a perturbation to the nonlinear equilibrium solution, a linearized equilibrium solution can be computed. We have demonstrated this technique for computing the shape gradient of the magnetic well for axisymmetric equilibria \citep{Paul2020b}. In generalizing this approach to 3D equilibria, there are additional challenges that arise. As regular singular points occur at every surface where the rotational transform resonates with a mode included in the spectrum for the solution vector, additional care must be taken in regularizing the equations. To avoid these difficulties associated with 3D MHD equilibria, the adjoint equations for a vacuum or force-free equilibrium model could instead be considered. 

We have presented several example configurations obtained with the set of figures of merit for which derivative information is available. Upon further advancement of adjoint methods, we hope that these numerical advances will enable the identification of equilibria of experimental relevance. To conclude, we anticipate many extensions of this work, such as applying the same adjoint principle for free-boundary optimization.

\section*{Acknowledgements}

The authors would like to acknowledge C. Zhu for the development of a python interface with the VMEC code and S. R. Hudson for enlightening discussions and help with the SPEC calculations. This work was supported by the US Department of Energy through grants DE-FG02-93ER-54197 and DE-FC02-08ER-54964. This work was also supported by a grant from the Simons Foundation (560651, ML). Some of the computations presented in this paper have used resources at the National Energy Research Scientific Computing Center (NERSC). 

\appendix

\section{Target rotational transform optimization}
\label{app:target_iota}

 In this Section, we perform a benchmark to demonstrate the convergence of the gradient-based optimization to a known minimum in a 2D space using the following objective function,
\begin{align}
  \widetilde{f}_{\iota} =  \frac{1}{2}\left(f_{\iota} - 0.06\right)^2,
  \label{eq:iota_target_benchmark}
\end{align}
defined with $\iota_{\text{target}} = 0.618034$ with the definition in \eqref{eq:iota_target}. Although this target function is not physically relevant, as there remains residual error in the rotational transform profile, it allows us to easily identify a local optimum where $f_{\iota} = 0.06$.

\begin{figure}
    \centering
    \begin{subfigure}[b]{0.49\textwidth}
    \includegraphics[width=1.0\textwidth]{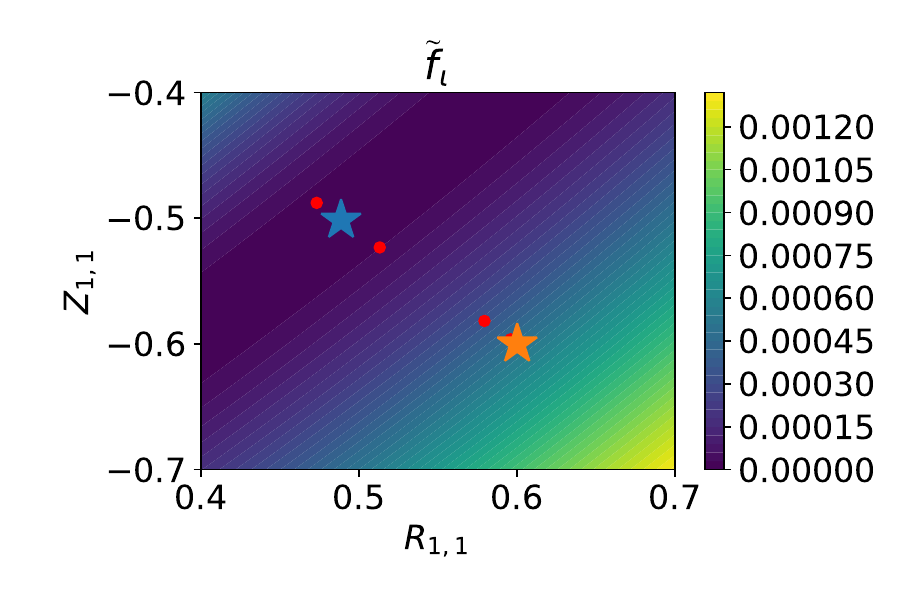}
    \caption{}
    \end{subfigure}
    \begin{subfigure}[b]{0.49\textwidth}
    \includegraphics[width=0.9\textwidth]{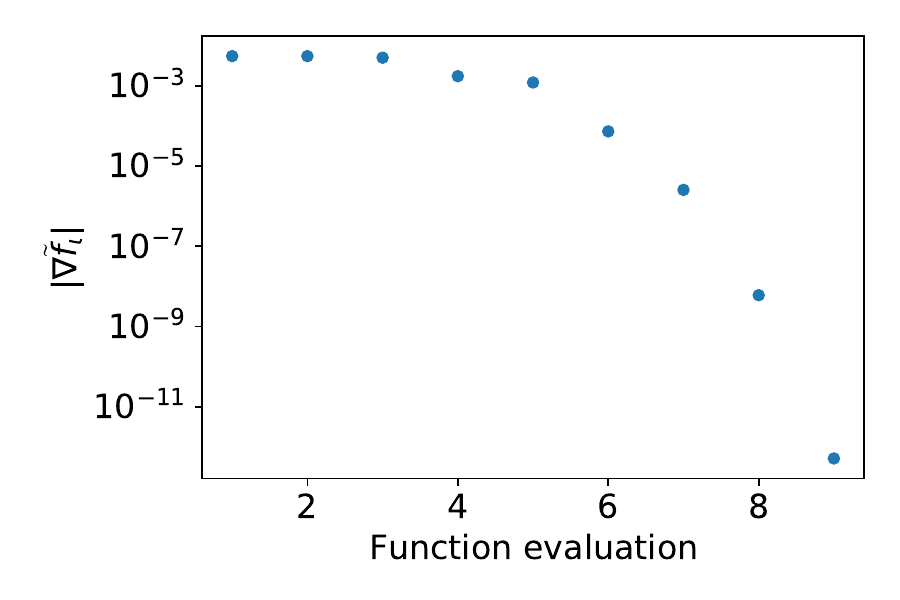}
    \caption{}
    \end{subfigure}
    \caption{(a) Convergence of BFGS optimization beginning at the orange star toward optimum function value at $R_{1,1} = 0.46$ and $Z_{1,1} = -0.47$ (blue star). Each function evaluation is denoted by a red dot. (b) The convergence of the L2 norm of the gradient of the objective function \eqref{eq:iota_target_benchmark}.}
    \label{fig:iota_2d}
\end{figure}

We begin with the boundary,
\begin{subequations}
\begin{align}
    R(\theta,\phi) &= R_{0} + R_{1,0}\cos(\theta) + R_{1,1}\cos(\theta - N_P \phi) \\
    Z(\theta,\phi) &= Z_{1,0}\sin(\theta) + Z_{1,1} \sin(\theta - N_P \phi),
\end{align}
\label{eq:rotating_ellipse}
\end{subequations}
with $R_0 = 5$, $R_{1,0} = 1.5$ , $R_{1,1} = 0.6$, $Z_{1,0} = 1.5$, and $Z_{1,1} = -0.6$. We consider a vacuum field with profiles $p(\psi) = I_T(\psi) = 0$. We identify the minimum of $\widetilde{f}_{\iota}$ with respect to $R_{1,1}$ and $Z_{1,1}$ by performing a scan over the local space. We achieve convergence to this optimum in 9 function evaluations (6 BFGS iterations). In Figure \ref{fig:iota_2d} we present the convergence of the objective function in the 2D optimization space, with the optimum denoted by the blue star. We are able to reduce the L2 gradient norm to $5.13 \times 10^{-13}$. 

\vspace{1cm}

\bibliographystyle{jpp}
\bibliography{bibliography.bib}

\begin{thebibliography}{53}
\expandafter\ifx\csname natexlab\endcsname\relax\def\natexlab#1{#1}\fi
\def\au#1{#1} \def\ed#1{#1} \def\yr#1{#1}\def\at#1{#1}\def\jt#1{\textit{#1}}
  \def\bt#1{#1}\def\bvol#1{\textbf{#1}} \def\vol#1{#1} \def\pg#1{#1}
  \def\publ#1{#1}\def\arxiv#1{#1}\def\org#1{#1}\def\st#1{\textit{#1}}

\bibitem[Anderson {\em et~al.\/}(1995)Anderson, Almagri, Anderson, Matthews,
  Talmadge \& Shohet]{Anderson1995}
{\sc \au{Anderson, F. S.~B.}, \au{Almagri, A.~F.}, \au{Anderson, D.~T.},
  \au{Matthews, P.~G.}, \au{Talmadge, J.~N.} \& \au{Shohet, J.~L.}} \yr{1995}
  \at{The {Helically Symmetric eXperiment},{(HSX)} goals, design and status}.
  \jt{Fusion Technology}  \bvol{27}~(3T),  \pg{273--277}.

\bibitem[Antonsen {\em et~al.\/}(2019)Antonsen, Paul \&
  Landreman]{Antonsen2019}
{\sc \au{Antonsen, T.}, \au{Paul, E.~J.} \& \au{Landreman, M.}} \yr{2019}
  \at{Adjoint approach to calculating shape gradients for three-dimensional
  magnetic confinement equilibria}.  \jt{Journal of Plasma Physics}
  \bvol{85}~(2).

\bibitem[Beidler {\em et~al.\/}(1990)Beidler, Grieger, Herrnegger, Harmeyer,
  Kisslinger, Lotz, Maassberg, Merkel, N{\"u}hrenberg, Rau {\em
  et~al.\/}]{Beidler1990}
{\sc \au{Beidler, C.}, \au{Grieger, G.}, \au{Herrnegger, F.}, \au{Harmeyer,
  E.}, \au{Kisslinger, J.}, \au{Lotz, W.}, \au{Maassberg, H.}, \au{Merkel, P.},
  \au{N{\"u}hrenberg, J.}, \au{Rau, F.} \& \au{others}} \yr{1990}  \at{Physics
  and engineering design for {Wendelstein VII-X}}.  \jt{Fusion Technology}
  \bvol{17}~(1),  \pg{148--168}.

\bibitem[Carlen {\em et~al.\/}(2005)Carlen, Laurie, Maddocks \&
  Smutny]{Carlen2005}
{\sc \au{Carlen, M.}, \au{Laurie, B.}, \au{Maddocks, J.~H.} \& \au{Smutny, J.}}
  \yr{2005}  \at{Biarcs, global radius of curvature, and the computation of
  ideal knot shapes}.  \bt{In {\em Physical and Numerical Models in Knot
  Theory: Including Applications to the Life Sciences\/}},  \pg{pp. 75--108}.
  \publ{World Scientific}.

\bibitem[Carlton-Jones {\em et~al.\/}(2020)Carlton-Jones, Paul \&
  Dorland]{Carlton2020}
{\sc \au{Carlton-Jones, A.}, \au{Paul, E.~J.} \& \au{Dorland, W.}} \yr{2020}
  \at{Computing the shape gradient of stellarator coil complexity with respect
  to the plasma boundary}.  \jt{arXiv preprint arXiv:2011.03702} .

\bibitem[Cooper {\em et~al.\/}(1992)Cooper, Hirshman, Merazzi \&
  Gruber]{Cooper19923d}
{\sc \au{Cooper, W.~A.}, \au{Hirshman, S.~P.}, \au{Merazzi, S.} \& \au{Gruber,
  R.}} \yr{1992}  \at{{3D} magnetohydrodynamic equilibria with anisotropic
  pressure}.  \jt{Computer Physics Communications}  \bvol{72}~(1),  \pg{1}.

\bibitem[Dekeyser(2014)]{Dekeyser2014}
{\sc \au{Dekeyser, W.}} \yr{2014}  \at{Optimal plasma edge configurations for
  next-step fusion reactors}. PhD thesis, Katholieke Universiteit Leuven.

\bibitem[Drevlak {\em et~al.\/}(2018)Drevlak, Beidler, Geiger, Helander \&
  Turkin]{Drevlak2018}
{\sc \au{Drevlak, M.}, \au{Beidler, C.~D.}, \au{Geiger, J.}, \au{Helander, P.}
  \& \au{Turkin, Y.}} \yr{2018}  \at{Optimisation of stellarator equilibria
  with {ROSE}}.  \jt{Nuclear Fusion}  \bvol{59}~(1),  \pg{016010}.

\bibitem[Drevlak {\em et~al.\/}(2013)Drevlak, Brochard, Helander, Kisslinger,
  Mikhailov, N{\"u}hrenberg, N{\"u}hrenberg \& Turkin]{Drevlak2013}
{\sc \au{Drevlak, M.}, \au{Brochard, F.}, \au{Helander, P.}, \au{Kisslinger,
  J.}, \au{Mikhailov, M.}, \au{N{\"u}hrenberg, C.}, \au{N{\"u}hrenberg, J.} \&
  \au{Turkin, Y.}} \yr{2013}  \at{{ESTELL: A quasi-toroidally symmetric
  stellarator}}.  \jt{Contributions to Plasma Physics}  \bvol{53}~(6),
  \pg{459--468}.

\bibitem[Drevlak {\em et~al.\/}(2014)Drevlak, Geiger, Helander \&
  Turkin]{Drevlak2014}
{\sc \au{Drevlak, M.~C.}, \au{Geiger, J.}, \au{Helander, P.} \& \au{Turkin,
  Y.}} \yr{2014}  \at{Fast particle confinement with optimized coil currents in
  the {W7-X} stellarator}.  \jt{Nuclear Fusion}  \bvol{54}~(7),  \pg{073002}.

\bibitem[Emmerich {\em et~al.\/}(2002)Emmerich, Giotis, {\"O}zdemir, B{\"a}ck
  \& Giannakoglou]{Emmerich2002}
{\sc \au{Emmerich, Michael}, \au{Giotis, Alexios}, \au{{\"O}zdemir, Mutlu},
  \au{B{\"a}ck, Thomas} \& \au{Giannakoglou, Kyriakos}} \yr{2002}
  Metamodel—assisted evolution strategies.  \bt{In {\em International
  Conference on parallel problem solving from nature\/}},  \pg{pp. 361--370}.
  Springer.

\bibitem[Gelsey(1995)]{Gelsey1995}
{\sc \au{Gelsey, Andrew}} \yr{1995}  \at{Intelligent automated quality control
  for computational simulation}.  \jt{AI EDAM}  \bvol{9}~(5),  \pg{387--400}.

\bibitem[Gelsey {\em et~al.\/}(1998)Gelsey, Schwabacher \& Smith]{Gelsey1998}
{\sc \au{Gelsey, Andrew}, \au{Schwabacher, Mark} \& \au{Smith, Don}} \yr{1998}
  \at{Using modeling knowledge to guide design space search}.  \jt{Artificial
  Intelligence}  \bvol{101}~(1-2),  \pg{35--62}.

\bibitem[Giuliani {\em et~al.\/}(2020)Giuliani, Wechsung, Cerfon, Stadler \&
  Landreman]{Giuliani2020}
{\sc \au{Giuliani, A.}, \au{Wechsung, F.}, \au{Cerfon, A.}, \au{Stadler, G.} \&
  \au{Landreman, M.}} \yr{2020}  \at{Single-stage gradient-based stellarator
  coil design: Optimization for near-axis quasi-symmetry}.  \jt{arXiv preprint
  arXiv:2010.02033} .

\bibitem[Gonzalez \& Maddocks(1999)]{Gonzalez1999}
{\sc \au{Gonzalez, O.} \& \au{Maddocks, J.~H.}} \yr{1999}  \at{Global
  curvature, thickness, and the ideal shapes of knots}.  \jt{Proceedings of the
  National Academy of Sciences}  \bvol{96}~(9),  \pg{4769--4773}.

\bibitem[Greene {\em et~al.\/}(1986)Greene, MacKay \& Stark]{Greene1986}
{\sc \au{Greene, J.~M.}, \au{MacKay, R.~S.} \& \au{Stark, J.}} \yr{1986}
  \at{Boundary circles for area-preserving maps}.  \jt{Physica D: Nonlinear
  Phenomena}  \bvol{21}~(2-3),  \pg{267--295}.

\bibitem[Helander(2014)]{Helander2014}
{\sc \au{Helander, P.}} \yr{2014}  \at{Theory of plasma confinement in
  non-axisymmetric magnetic fields}.  \jt{Reports on Progress in Physics}
  \bvol{77}~(8),  \pg{087001}.

\bibitem[Henneberg {\em et~al.\/}(2019)Henneberg, Drevlak, N{\"u}hrenberg,
  Beidler, Turkin, Loizu \& Helander]{Henneberg2019}
{\sc \au{Henneberg, SA}, \au{Drevlak, M}, \au{N{\"u}hrenberg, C}, \au{Beidler,
  CD}, \au{Turkin, Y}, \au{Loizu, J} \& \au{Helander, P}} \yr{2019}
  \at{Properties of a new quasi-axisymmetric configuration}.  \jt{Nuclear
  Fusion}  \bvol{59}~(2),  \pg{026014}.

\bibitem[Hirshman \& Whitson(1983)]{Hirshman1983}
{\sc \au{Hirshman, S.~P.} \& \au{Whitson, J.~C.}} \yr{1983}
  \at{Steepest‐descent moment method for three‐dimensional
  magnetohydrodynamic equilibria}.  \jt{The Physics of Fluids}  \bvol{26}~(12),
   \pg{3553--3568}.

\bibitem[Hudson {\em et~al.\/}(2012)Hudson, Dewar, Dennis, Hole, McGann,
  Von~Nessi \& Lazerson]{Hudson2012}
{\sc \au{Hudson, S.~R.}, \au{Dewar, R.~L.}, \au{Dennis, G.}, \au{Hole, M.~J.},
  \au{McGann, M.}, \au{Von~Nessi, G.} \& \au{Lazerson, S.}} \yr{2012}
  \at{Computation of multi-region relaxed magnetohydrodynamic equilibria}.
  \jt{Physics of Plasmas}  \bvol{19}~(11),  \pg{112502}.

\bibitem[Johnson(2014)]{NLOPT}
{\sc \au{Johnson, S.~G.}} \yr{2014} The {NLopt} nonlinear-optimization package.

\bibitem[Kesner {\em et~al.\/}(1995)Kesner, Ramos \& Gang]{Kesner1995}
{\sc \au{Kesner, J.}, \au{Ramos, J.~J.} \& \au{Gang, F.-Y.}} \yr{1995}
  \at{Comet cross-section tokamaks}.  \jt{Journal of Fusion Energy}
  \bvol{14}~(4),  \pg{361--371}.

\bibitem[Landreman(2020)]{Landreman2020a}
{\sc \au{Landreman, M.}} \yr{2020} Figures of merit for stellarators near the
  magnetic axis,  \arxiv{arXiv: 2012.00865}.

\bibitem[Landreman \& Jorge(2020)]{Landreman2020b}
{\sc \au{Landreman, M.} \& \au{Jorge, R.}} \yr{2020}  \at{Magnetic well and
  {Mercier} stability of stellarators near the magnetic axis}.  \jt{Journal of
  Plasma Physics}  \bvol{86}~(5),  \pg{905860510}.

\bibitem[Landreman \& Sengupta(2019)]{Landreman2019}
{\sc \au{Landreman, M.} \& \au{Sengupta, W.}} \yr{2019}  \at{Constructing
  stellarators with quasisymmetry to high order}.  \jt{Journal of Plasma
  Physics}  \bvol{85}~(6).

\bibitem[Lazerson {\em et~al.\/}(2020)Lazerson, Schmitt, Zhu, Breslau \&
  {STELLOPT Developers}]{Lazerson2020}
{\sc \au{Lazerson, Samuel}, \au{Schmitt, John}, \au{Zhu, Caoxiang},
  \au{Breslau, Joshua} \& \au{{STELLOPT Developers}}} \yr{2020} {STELLOPT}.
  https://github.com/PrincetonUniversity/STELLOPT.

\bibitem[Lemar{\'e}chal(1982)]{Lemarechal1982}
{\sc \au{Lemar{\'e}chal, Claude}} \yr{1982}  \at{Numerical experiments in
  nonsmooth optimization.} .

\bibitem[Lewis \& Overton(2009)]{Lewis2009}
{\sc \au{Lewis, Adrian~S} \& \au{Overton, Michael~L}} \yr{2009}  \at{Nonsmooth
  optimization via {BFGS}}.  \jt{Submitted to SIAM J. Optimiz}  \pg{pp. 1--35}.

\bibitem[McGreivy {\em et~al.\/}(2021)McGreivy, Hudson \& Zhu]{Mcgreivy2021}
{\sc \au{McGreivy, N.}, \au{Hudson, S.R.} \& \au{Zhu, C.}} \yr{2021}
  \at{Optimized finite-build stellarator coils using automatic
  differentiation}.  \jt{Nuclear Fusion}  \bvol{61}~(2),  \pg{026020}.

\bibitem[Medvedev {\em et~al.\/}(2015)Medvedev, Kikuchi, Villard, Takizuka,
  Diamond, Zushi, Nagasaki, Duan, Wu, Ivanov {\em et~al.\/}]{Medvedev2015}
{\sc \au{Medvedev, S.~Y.}, \au{Kikuchi, M.}, \au{Villard, L.}, \au{Takizuka,
  T.}, \au{Diamond, P.}, \au{Zushi, H.}, \au{Nagasaki, K.}, \au{Duan, X.},
  \au{Wu, Y.}, \au{Ivanov, A.~A.} \& \au{others}} \yr{2015}  \at{The negative
  triangularity tokamak: stability limits and prospects as a fusion energy
  system}.  \jt{Nuclear Fusion}  \bvol{55}~(6),  \pg{063013}.

\bibitem[Meiss(1992)]{Meiss1992}
{\sc \au{Meiss, J.~D.}} \yr{1992}  \at{Symplectic maps, variational principles,
  and transport}.  \jt{Reviews of Modern Physics}  \bvol{64}~(3),  \pg{795}.

\bibitem[Mercier(1964)]{Mercier1964}
{\sc \au{Mercier, C.}} \yr{1964}  \at{Equilibrium and stability of a toroidal
  magnetohydrodynamic system in the neighbourhood of a magnetic axis}.
  \jt{Nuclear Fusion}  \bvol{4}~(3),  \pg{213}.

\bibitem[Mercier \& Luc(1974)]{Mercier1974}
{\sc \au{Mercier, C.} \& \au{Luc, H.}} \yr{1974}  \at{The {MHD} approach to the
  problem of plasma confinement in closed magnetic configurations}.
  \jt{Lectures in Plasma Physics, Commission of the European Communities,
  Luxembourg} .

\bibitem[Miner~Jr {\em et~al.\/}(2001)Miner~Jr, Valanju, Hirshman, Brooks \&
  Pomphrey]{Miner2001}
{\sc \au{Miner~Jr, W.~H.}, \au{Valanju, P.~M.}, \au{Hirshman, S.~P.},
  \au{Brooks, A.} \& \au{Pomphrey, N.}} \yr{2001}  \at{Use of a genetic
  algorithm for compact stellarator coil design}.  \jt{Nuclear Fusion}
  \bvol{41}~(9),  \pg{1185}.

\bibitem[Mynick {\em et~al.\/}(2002)Mynick, Pomphrey \& Ethier]{Mynick2002}
{\sc \au{Mynick, H.~E.}, \au{Pomphrey, N.} \& \au{Ethier, S.}} \yr{2002}
  \at{Exploration of stellarator configuration space with global search
  methods}.  \jt{Physics of Plasmas}  \bvol{9}~(3),  \pg{869--876}.

\bibitem[Nocedal \& Wright(2006)]{Nocedal2006}
{\sc \au{Nocedal, J.} \& \au{Wright, S.~J.}} \yr{2006} {\em Numerical
  Optimization\/}.  \publ{Springer}.

\bibitem[N{\"u}hrenberg \& Zille(1988)]{Nuhrenberg1988}
{\sc \au{N{\"u}hrenberg, J.} \& \au{Zille, R.}} \yr{1988}  \at{Quasi-helically
  symmetric toroidal stellarators}.  \jt{Physics Letters A}  \bvol{129}~(2),
  \pg{113--117}.

\bibitem[Oymak \& Soltanolkotabi(2018)]{Oymak2018}
{\sc \au{Oymak, S.} \& \au{Soltanolkotabi, M.}} \yr{2018}
  \at{Overparameterized nonlinear learning: Gradient descent takes the shortest
  path?}  \jt{arXiv preprint arXiv:1812.10004} .

\bibitem[Paul(2020)]{Paul2020b}
{\sc \au{Paul, E.~J.}} \yr{2020}  \at{Adjoint methods for stellarator shape
  optimization and sensitivity analysis}.  \jt{arXiv preprint arXiv:2005.07633}
  .

\bibitem[Paul {\em et~al.\/}(2019)Paul, Abel, Landreman \& Dorland]{Paul2019}
{\sc \au{Paul, E.~J.}, \au{Abel, I.~G.}, \au{Landreman, M.} \& \au{Dorland,
  W.}} \yr{2019}  \at{An adjoint method for neoclassical stellarator
  optimization}.  \jt{Journal of Plasma Physics}  \bvol{85}~(5).

\bibitem[Paul {\em et~al.\/}(2020)Paul, Antonsen, Landreman \&
  Cooper]{Paul2020}
{\sc \au{Paul, E.~J.}, \au{Antonsen, T.}, \au{Landreman, M.} \& \au{Cooper,
  W.~A.}} \yr{2020}  \at{Adjoint approach to calculating shape gradients for
  three-dimensional magnetic confinement equilibria. {Part 2. Applications}}.
  \jt{Journal of Plasma Physics}  \bvol{86}~(1).

\bibitem[Paul {\em et~al.\/}(2018)Paul, Landreman, Bader \& Dorland]{Paul2018}
{\sc \au{Paul, E.~J.}, \au{Landreman, M.}, \au{Bader, A.} \& \au{Dorland, W.}}
  \yr{2018}  \at{An adjoint method for gradient-based optimization of
  stellarator coil shapes}.  \jt{Nuclear Fusion}  \bvol{58}~(7),  \pg{076015}.

\bibitem[Pogutse \& Yurchenko(1982)]{Pogutse1982}
{\sc \au{Pogutse, O.} \& \au{Yurchenko, E.}} \yr{1982}  \at{{Reviews of Plasma
  Physics}}.  \jt{Consultants Bureau, New York} .

\bibitem[Rasheed {\em et~al.\/}(1997)Rasheed, Hirsh \& Gelsey]{Rasheed1997}
{\sc \au{Rasheed, Khaled}, \au{Hirsh, Haym} \& \au{Gelsey, Andrew}} \yr{1997}
  \at{A genetic algorithm for continuous design space search}.  \jt{Artificial
  Intelligence in Engineering}  \bvol{11}~(3),  \pg{295--305}.

\bibitem[Rodriguez {\em et~al.\/}(2020)Rodriguez, Helander \&
  Bhattacharjee]{Rodriguez2020}
{\sc \au{Rodriguez, E.}, \au{Helander, P.} \& \au{Bhattacharjee, A.}} \yr{2020}
   \at{Necessary and sufficient conditions for quasisymmetry}.  \jt{Physics of
  Plasmas}  \bvol{27}~(6),  \pg{062501}.

\bibitem[Sanchez {\em et~al.\/}(2000)Sanchez, Hirshman, Ware, Berry \&
  Spong]{Sanchez2000}
{\sc \au{Sanchez, R.}, \au{Hirshman, S.~P.}, \au{Ware, A.~S.}, \au{Berry,
  L.~A.} \& \au{Spong, D.~A.}} \yr{2000}  \at{Ballooning stability optimization
  of low-aspect-ratio stellarators}.  \jt{Plasma Physics and Controlled Fusion}
   \bvol{42}~(6),  \pg{641}.

\bibitem[Sauer(2012)]{Sauer2012}
{\sc \au{Sauer, Timothy}} \yr{2012} {\em Numerical Analysis\/}.
  \publ{Pearson}.

\bibitem[Smutny(2004)]{Smutny2004}
{\sc \au{Smutny, J.}} \yr{2004}  \at{Global radii of curvature, and the biarc
  approximation of space curves: In pursuit of ideal knot shapes}. PhD thesis,
  EPFL.

\bibitem[Taylor(1965)]{Taylor1965}
{\sc \au{Taylor, J.~B.}} \yr{1965}  \at{Simple toroidal magnetic field with
  negative {$V''$}}.  \jt{The Physics of Fluids}  \bvol{8}~(6),
  \pg{1203--1205}.

\bibitem[Walker(2016)]{Walker2016}
{\sc \au{Walker, S.~W.}} \yr{2016}  \at{Shape optimization of self-avoiding
  curves}.  \jt{Journal of Computational Physics}  \bvol{311},  \pg{275--298}.

\bibitem[Wesson \& Campbell(2011)]{Wesson2011}
{\sc \au{Wesson, J.} \& \au{Campbell, D.~J.}} \yr{2011} {\em Tokamaks\/},
  \st{International Series of Monographs on Physics},  \vol{vol. 149}.
  \publ{Oxford University Press}.

\bibitem[Zarnstorff {\em et~al.\/}(2001)Zarnstorff, Berry, Brooks, Fredrickson,
  Fu, Hirshman, Hudson, Ku, Lazarus, Mikkelsen {\em et~al.\/}]{Zarnstorff2001}
{\sc \au{Zarnstorff, M.~C.}, \au{Berry, L.~A.}, \au{Brooks, A.},
  \au{Fredrickson, E.}, \au{Fu, G.~Y.}, \au{Hirshman, S.}, \au{Hudson, S.},
  \au{Ku, L.~P.}, \au{Lazarus, E.}, \au{Mikkelsen, D.} \& \au{others}}
  \yr{2001}  \at{Physics of the compact advanced stellarator {NCSX}}.
  \jt{Plasma Physics and Controlled Fusion}  \bvol{43}~(12A),  \pg{A237}.

\bibitem[Zhu {\em et~al.\/}(2018)Zhu, Hudson, Song \& Wan]{Zhu2018}
{\sc \au{Zhu, C.}, \au{Hudson, S.~R.}, \au{Song, Y.} \& \au{Wan, Y.}} \yr{2018}
   \at{{New method to design stellarator coils without the winding surface}}.
  \jt{Nuclear Fusion}  \bvol{58},  \pg{016008}.

\end{thebibliography}

\end{document}